\documentclass[twocolumn,lineno,dvipsnames, usenames]{aastex62}
\usepackage{amssymb,amsmath,verbatim,mathtools,needspace,enumitem,etoolbox,graphicx,physics,microtype,ulem}
\usepackage{xspace}
\definecolor{linkcolor}{rgb}{0.0,0.3,0.5}
\definecolor{urlcolor}{rgb}{0.27,0.55,0.}
\definecolor{funcolor}{rgb}{0.65, 0.16, 0.16}
\usepackage[all]{hypcap}
\usepackage{slashbox,booktabs,amsmath}

\newcommand{\Eiso}{E^{\mathrm{iso}}}
\newcommand{\Mtot}{M_{\mathrm{tot}}}
\newcommand{\Msun}{M_{\odot}}
\newcommand{\LIGOlabMIT}{\affiliation{LIGO Laboratory, Massachusetts Institute of Technology, 185 Albany St, Cambridge, MA 02139, USA}}
\newcommand{\MKI}{\affiliation{Department of Physics and Kavli Institute for Astrophysics and Space Research, Massachusetts Institute of Technology, 77 Massachusetts Ave, Cambridge, MA 02139, USA}}
\newcommand{\Monash}{\affiliation{School of Physics and Astronomy, Monash University, Clayton, VIC 3800, Australia}}
\newcommand{\Ozgrav}{\affiliation{OzGrav: The ARC Centre of Excellence for Gravitational-Wave Discovery, Clayton, VIC 3800, Australia}}
\begin{document}

\title{Constraining short gamma-ray burst jet properties with gravitational waves and gamma rays}
\author{Sylvia Biscoveanu} \LIGOlabMIT \MKI \Monash \Ozgrav
\email{sbisco@mit.edu}
\author{Eric Thrane} \Monash \Ozgrav
\author{Salvatore Vitale} \LIGOlabMIT \MKI

\begin{abstract}
Gamma-ray burst (GRB) prompt emission is highly beamed, and understanding the jet geometry and beaming configuration can provide information on the poorly understood central engine and circum-burst environment.
Prior to the advent of gravitational-wave astronomy, astronomers relied on observations of jet breaks in the multi-wavelength afterglow to determine the GRB opening angle, since the observer's viewing angle relative to the system cannot be determined from the electromagnetic data alone.
Gravitational-wave observations, however, provide an independent measurement of the viewing angle.
We describe a Bayesian method for determining the geometry of short GRBs using coincident electromagnetic and gravitational-wave observations.
We demonstrate how an ensemble of multi-messenger detections can be used to measure the distributions of the jet energy, opening angle, Lorentz factor, and angular profile of short GRBs; we find that for a population of 100 such observations, we can constrain the mean of the opening angle distribution to within $10^{\circ}$ regardless of the angular emission profile. Conversely, the constraint on the energy distribution depends on the shape of the profile, which can be distinguished.
\end{abstract}
\keywords{Gravitational waves, neutron stars, compact binaries, gamma-ray bursts}

\section{Introduction}
Understanding the emission profile and jet geometry of gamma-ray bursts (GRBs) has wide-ranging implications for the energetics, rates, and luminosity function of these relativistic explosions, all of which ultimately provide insight into the nature of the central engine. The GRB population is bimodal in duration and hardness, with long-soft and short-hard bursts defined by a transition at $\sim 2~\mathrm{s}$~\citep{bimodal}. %
Short GRB (sGRB) afterglows are uniformly fainter than their long GRB counterparts (e.g.~\cite{sgrb_afterglow_faint,Fong_sgrb_catalog}, see~\cite{nakar_review, berger_review} for reviews) and were first observed in 2005~\citep{first_sgrb_afterglow_hjorth, first_sgrb_afterglow_fox, first_sgrb_afterglow_gehrels, first_sgrb_afterglow_villasenor,first_sgrb_afterglow_barthelmy}. The lack of an associated supernova (e.g.~\cite{first_sgrb_afterglow_fox,kocevski_nosn,sgrb_jet_break_soderberg,davanzo_nosn}) together with the localization of some short GRBs to early-type galaxies~\citep{fong_berger_sgrb_hosts,fong_etal_sgrb_hosts,prochaska_galaxy_type} provided early evidence in support of the binary neutron star or neutron star-black hole merger progenitor model~\citep{eichler_bns_progenitors,narayan_bns_progenitors}. The recent coincident detection~\citep{gw170817_grb_association} of gravitational-wave event GW170817 from a binary neutron star merger~\citep{gw170817} and the short, hard burst GRB 170817A~\citep{grb170817A_fermi,grb170817A_integral} has confirmed the compact binary progenitor model for at least some sGRBs.

In this paper, we describe a Bayesian method to combine gravitational-wave and electromagnetic observations of sGRBs from binary neutron star coalescences to infer the total energy and Lorentz factor of the jet as well as the opening angle and power law index of the jet emission profile. 
Because gravitational-wave data provides an independent measurement of the inclination angle between the jet axis and the observer's line of sight\footnote{The gravitational-wave signal provides an independent measurement of the angles between the binary angular momentum and the line of sight. Based on the results of fully general relativistic magnetohydrodynamical simulations, the jet is believed to be emitted along the spin axis of the remnant black hole due to the presence of a strong poloidal magnetic field, so the viewing angle of the GRB is expected to coincide with the inclination angle measured with gravitational waves~\citep{Rezzolla_mass1, rezzolla_mass2, Giazomazzo_mass}.}, the opening angle can be inferred directly from the prompt emission, eliminating the dependence on afterglow observations imposed by the traditional jet break calculation. 

Our analysis builds on previous Bayesian methods for combining GW and EM data, which have been used to provide improved estimates of the neutron star parameters like the mass ratio and tidal deformability for GW170817~\citep{Coughlin_2018, Radice_2018, Capano_2019, Coughlin_2019, Radice_2019}.

\cite{glasgow_bayes1,glasgow_bayes2} have previously shown that combining gravitational-wave and GRB data can also be used to determine the GRB luminosity function and host galaxy and offer improved inference of parameters already constrained by the gravitational-wave data alone like the inclination angle and distance to the source system. 
While previous studies have offered constraints on the jet opening angle using estimates of the coincident gravitational-wave/GRB detection rate for top-hat jets~\citep{chen_beaming, Clark_2015, glasgow_rates} and by fitting the GRB luminosity assuming a structured jet geometry in conjunction with estimates of the binary neutron star merger rate~\citep{Mogushi_2019}, we seek to measure the GRB energy, Lorentz factor, opening angle, and power law index directly by parameterizing the measured fluence of the GRB prompt emission in terms of these four parameters and additional parameters inferred from the GW data. We analyze a simulated population of coincident GW and GRB detections to determine what type of constraints can be derived on the distributions of these parameters by combining an ensemble of multiple coincident events. 

The rest of this paper is organized as follows. 
We first discuss the jet break method for estimating the jet opening angle and its applications to GRB 170817A in Section~\ref{sec:jet_breaks}, and then describe the top-hat and universal structured jet energy models, as well as the prescription for calculating the observed GRB fluence for a given jet geometry and inclination angle in Section~\ref{sec:energy_models}. In Section~\ref{sec:methods} we outline the Bayesian parameter estimation method that we use to combine the GW and EM measurements for individual events and the hierarchical model used to determine the population hyper-parameters. We present results for simulated top-hat and structured power-law jet populations in Section~\ref{sec:results}, and conclude in Section~\ref{sec:discussion} with a discussion of the implications of this study.

\section{Existing Observational Constraints}
\label{sec:jet_breaks}
The observational signature of collimated jets in GRBs is an achromatic jet break in the afterglow light curve that occurs at $t_{j}$ after the prompt emission, when the bulk Lorentz factor of the outflow has decreased to $\Gamma \approx 1/\theta_{j}$, where $\theta_{j}$ is the half-opening angle of the jet~\citep{rhoads_jet_break_1997,rhoads_jet_break_1999,sari_jet_break}.
The break is caused by a combination of two effects. The first is an edge effect that occurs when the entire emitting surface of the jet becomes visible. 
Due to relativistic beaming, the emission appears to come from a small fraction of the visible area, so the ``missing" component relative to the expectation from a spherical outflow manifests itself as a steepening in the light curve. Simultaneously, when the jet edge comes into causal contact with the  jet center as the Lorentz factor decreases, the jet begins to spread laterally, and the energy per solid angle decreases with time and radius, again resulting in a steepening of the light curve (see \cite{granot_jet_review} for a review of GRB jets). The jet break is observable from the X-ray to the radio bands, and the opening angle can be calculated via~\citep{sari_jet_break}:
\begin{align}
\theta_{j}\approx 9.5^\circ\bigg(\frac{t_{j,\mathrm{d}}}{1+z}\bigg)^{3/8}\bigg(\frac{n_{0}}{E_{\mathrm{K,iso,}52}}\bigg)^{1/8} ,
\end{align}
where $t_{j,\mathrm{d}}$ is the jet break time measured in days, $n_{0}$ is the density of the circumburst medium in $\mathrm{cm}^{-3}$, and $E_{\mathrm{K,iso,}52}$ is the isotropic equivalent kinetic energy of the ejecta in units of $10^{52}~\mathrm{erg}$.

Because the afterglow emission of most sGRBs decays much faster and at a uniform rate compared to long GRBs, jet breaks have only been reported for five sGRBs (GRBs 051221A~\citep{sgrb_jet_break_burrows,sgrb_jet_break_soderberg}, 090426A~\citep{sgrb_jet_break_guelbenzu}, 111020A~\citep{sgrb_jet_break_fong2}, 130603B~\citep{sgrb_jet_break_fong1}, and 140903A~\citep{sgrb_jet_break_troja}). Furthermore, the observation of a jet break provides no information on the structure of the jet. The two leading jet structure models that both predict a jet break in the afterglow light curve are the uniform, or \emph{top-hat jet}, where the energy per solid angle $\mathcal{E}$ and the Lorentz factor $\Gamma$ are constant over the entire emitting surface~\citep{rhoads_jet_break_1997,rhoads_jet_break_1999,sari_jet_break,granot_uj,panaitescu_uj}, and the \emph{universal structured jet}, where $\mathcal{E}$ and $\Gamma$ decay as a power law with the angle from the jet axis, $\theta^{-k}$~\citep{zhang_usj,rossi_usj}. While the jet break can be explained in terms of the intrinsic opening angle of the jet in the top-hat model, the universal structured jet model explains the jet break in terms of the viewing angle of the observer, implying that the opening angle of the jet is much wider than in the top-hat case. Other profiles, like a Gaussian structured jet or a radially stratified jet, can also reproduce the jet break behavior. All of these models make simplifying assumptions about the true angular emission profile, which would be obtained from hydrodynamical simulations in the ideal case where such simulations could reliably produce estimates of the GRB jet evolution.

The jet geometry of GRB 170817A~\citep{grb170817A_fermi,gw170817_grb_association} has been studied extensively. Its low luminosity together with the lack of \textit{early} X-ray~\citep{grb170817a_xray} and radio afterglow~\citep{grb170817a_xray} disfavors both of the simple top-hat and power-law universal structured jet models~\citep{grb170817a_kasliwal}, and is instead better explained by ``cocoon" emission; as the jet drills through the merger ejecta surrounding the central engine, it inflates a mildly relativistic cocoon. In addition to the internal shocks that arise in the jet, the interaction of the jet and the merger ejecta forms another set of forward and reverse shocks. The reverse shock heats the jet material and creates an inner cocoon surrounding the jet. The forward shock propagating into the merger ejecta forms the outer cocoon, which is only mildly relativistic, with Lorentz factors of a few~\citep{cocoon_gottlieb2,Lazzati_cocoon}. The forward shock continues to propagate through the ejecta as long as the medium is optically thick enough to sustain its width, at which point the radiation inside the shock layer breaks out, producing the observed $\gamma$-rays as the residual photons diffuse out of the cocoon (\cite{cocoon_gottlieb1,shock_Nakar_2010}; see also the shock breakout model of \cite{shock_Beloborodov}).

If the initial jet has a very short duration, low energy, or wide opening angle, it may be ``choked" by the cocoon. In this scenario, the jet does not manage to escape from the merger ejecta, and all of the initial energy of the jet is deposited into the cocoon. The observed $\gamma$-rays come entirely from the cocoon fireball~\citep{piran_fireball}.  If the initial jet launched by the central engine does manage to escape, it will still inflate a cocoon, so the observed $\gamma$-ray emission will consist of an ultra-relativistic, narrow core in addition to mildly relativistic cocoon ``wings"~\citep{cocoon_gottlieb1}. In this sense, the cocoon model provides physical motivation for the universal structured jet model.

In the case of GRB 170817A, the jet is under-luminous by several orders of magnitude compared to $E_{\mathrm{iso}}$ measurements for the rest of the short GRB population~\citep{gw170817_grb_association}. Light curve modeling revealed that it is impossible to reproduce the observed emission using a top-hat jet model viewed off axis for physically realistic values of the circum-burst density~\citep{grb170817a_kasliwal}. Instead, emission from a wide-angled cocoon that fades on the order of a few hours can explain both the prompt emission and the lack of early observations in the X-ray and radio bands that would be expected from the afterglow emission of a standard top-hat jet~\citep{grb170817a_radio, grb170817a_xray, cocoon_mooley, cocoon_gottlieb1}. It was impossible to determine if the gamma-ray emission from the cocoon was accompanied by a successful jet from the observed prompt emission alone. Follow-up radio observations using very long-baseline interferometry determined that the jet exhibited superluminal motion, indicating that a collimated jet with opening angle $\theta_{j} < 5~\mathrm{deg}$ successfully broke out of the cocoon~\citep{170817_vlbi}. This picture was confirmed by further analyses covering the entire afterglow spectrum~\citep{Margutti_2018, Lamb_2019, Ghirlanda_2019, ziaeepour2019}.

\section{GRB Energy Models}
\label{sec:energy_models}

\subsection{Isotropic Equivalent Energy}
The isotropic equivalent energy of the GRB in the source frame, $E^{\mathrm{iso}}$, is calculated by assuming the gamma-ray fluence, $F^{\gamma}$, measured by the observer is the same in all directions: $\Eiso = 4\pi F^{\gamma}d_{L}^{2}(1+z)^{-1}$, where $d_{L}$ is the luminosity distance and $z$ is the redshift of the source. The measured fluence and thus the isotropic equivalent energy depend on the observer's inclination angle, the total energy of the jet, and the emission profile.  
Consider a jet with fixed, uniform Lorentz factor $\Gamma$, that emits total energy $E^{\gamma}_{0}/2$ in the source frame\footnote{The source frame and the observer frame are the same except for the redshift correction.}. Each jet element is moving radially in the direction $\hat{n}$ at angle $\theta = \arccos{(\hat{n}\cdot \hat{z})}$ from the jet axis $\hat{z}$ and emits isotropically. The energy per solid angle emitted in the rest frame of each element is then:
\begin{align}
\mathcal{E}_{R}(\theta) \equiv \frac{dE}{d\Omega_{R}} = \frac{E^{\gamma}_{0}/2}{4\pi\Gamma}f_{R}(\theta)
\label{eq:epsilon_rest}
\end{align}
The profile function $f_{R}(\theta)$ is azimuthally symmetric around the jet axis, and is normalized so that
\begin{align}
\oint f_{R}(\theta)d\Omega_{R} = 2\pi\int^{\pi/2}_{0}f_{R}(\theta)\sin{\theta}d\theta = 1.
\end{align}
\begin{figure}
\centering
\centerline{\includegraphics[width=0.5\textwidth]{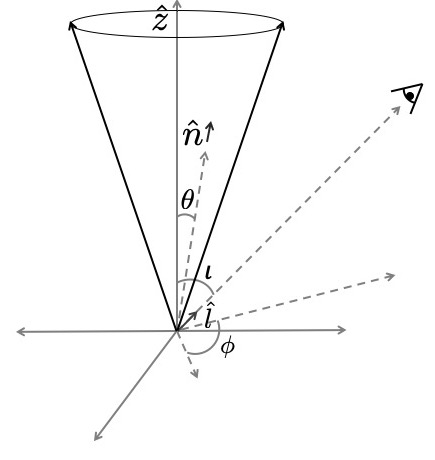}}
\caption{Coordinate system for the jet seen by an observer in the $\hat{l}$ direction. The angles $\theta$ and $\iota$ are defined such that $\hat{l}\cdot \hat{z} = \cos{\iota}$ and $\hat{n}\cdot \hat{z} = \cos{\theta}$ and $\phi$ is the angle between the projections of $\hat{l}$ and $\hat{n}$ in the $x-y$ plane.}
\label{fig:schematic}
\end{figure}

The profile function is defined in the rest frame and determines the brightness of each jet element as a function of its angular distance from the jet axis. The emission from each element, while isotropic in the rest frame, will appear highly beamed into a cone of angular width $\sim 1/\Gamma$ in the source frame due to relativistic beaming. The source-frame energy per solid angle can be calculated by applying the relativistic Doppler shift to Eq.~\ref{eq:epsilon_rest}~\citep{eichler_eiso_calc, graziani,salafia_eiso_calc}:
\begin{align}
&\mathcal{E}_{S}(\theta) = \oint \mathcal{E}_{R}(\theta)\Gamma^{-3}[1-\beta(\hat{l}\cdot \hat{n})]^{-3}d\Omega_{R}
\end{align}
where $\beta = v/c$ is the speed of the merger ejecta, and the Doppler factor is applied once for the energy and once for each angular dimension. The integral sums the Doppler-boosted contribution to the total energy from each jet element at angle $\theta$ relative to the jet axis. The inclination angle $\iota$ of the observer is encoded in the dot product, $\hat{l} \cdot \hat{n}= \cos{\theta}\cos{\iota}+\sin{\theta}\sin{\iota}\cos{\phi},$ between the unit vector in the direction of the jet element, $\hat{n}$, and the unit vector pointing towards the observer, $\hat{l}$, as illustrated in Fig.~\ref{fig:schematic}. The isotropic equivalent energy is calculated by assuming that the energy per unit solid angle measured at some inclination angle, $\iota$ is the same in all directions:
\begin{align}
&\Eiso(\iota) = 4\pi \mathcal{E}_{S}(\theta) \nonumber \\
&=\frac{E^{\gamma}_{0}}{2\Gamma^{4}}\int_{0}^{2\pi} \int_{0}^{\pi/2} \frac{f_{R}(\theta)\sin{\theta}\, d\theta\, d\phi}{[1-\beta(\cos{\theta}\cos{\iota}+\sin{\theta}\sin{\iota}\cos{\phi})]^{3}},
\end{align}
where we have substituted the definition of $\mathcal{E}_{R}(\theta)$ given in Eq.~\ref{eq:epsilon_rest}. If the Lorentz factor is also allowed to depend on the angle from the jet axis, the fluence at a particular inclination angle then becomes
\begin{align}\label{eq:fluence}
&F^{\gamma}(\iota) = \frac{E^{\gamma}_{0}(1+z)}{8\pi d_{L}^{2}}\times\\
&\int_{0}^{2\pi} \int_{0}^{\pi/2} \frac{f_{R}(\theta)\sin{\theta}\, d\theta\, d\phi}{\Gamma^{4}(\theta)[1-\beta(\theta)(\cos{\theta}\cos{\iota}+\sin{\theta}\sin{\iota}\cos{\phi})]^{3}}. \nonumber
\end{align}
\newline

\subsection{Uniform Jet Model}
Under the uniform jet or top-hat model, the energy per solid angle in the rest frame is expected to be constant within some jet opening angle $\theta_{j}$~\citep{granot_uj,panaitescu_uj}:
\begin{align}
\frac{dE}{d\Omega_{R}} = \mathcal{E}_{R}(\theta) =
\begin{cases}
\mathcal{E}_{0}, & \theta\leq \theta_{j}\\
0, &\theta > \theta_{j}
\end{cases}.
\end{align}

The total gamma-ray energy emitted in the rest frame can then be calculated by integrating over all solid angles and multiplying by a factor of 2 to account for both jets:
\begin{align}
E^{\gamma}_{0} &= 2\int_{0}^{2\pi}\int_{0}^{\theta_{j}}\mathcal{E}_{0}\sin{\theta '}d\theta '\, d\phi \nonumber\\
&=4\pi\mathcal{E}_{0}(1-\cos{\theta_{j}})\nonumber\\
&=4\pi \frac{dE}{d\Omega_{R}}f_{b},
\end{align}
where we have defined the beaming factor $f_{b} \equiv (1-\cos{\theta_{j}})$. For the uniform jet model, we recover the typical relationship between the total energy and the isotropic equivalent energy~\citep{long_grb_beaming}:
\begin{align}
\Eiso(\theta_{j}) \approx E_{0}^{\gamma}/f_{b}.
\end{align}

\subsection{Universal Structured Jet Model}
The universal structured jet model assumes that all GRBs have a quasi-universal beaming configuration and that the variability in jet break time is due to the inclination angle rather than the intrinsic opening angle of the jet itself. Both the energy per solid angle and the Lorentz factor fall off as power laws as a function of the angle from the jet axis~\citep{zhang_usj,rossi_usj}:
\begin{align}
\frac{dE}{d\Omega_{R}} = \mathcal{E}_{R}(\theta) =
\begin{cases}
\mathcal{E}_{0}, & \theta\leq \theta_{c}\\
\mathcal{E}_{0}(\theta/\theta_{c})^{-k}, & \theta_{c}<\theta\leq \theta_{j}\\
0, &\theta > \theta_{j},
\end{cases}
\\
\Gamma(\theta) =
\begin{cases}
\Gamma_{0}, & \theta\leq \theta_{c}\\
\Gamma_{0}(\theta/\theta_{c})^{-k}, & \theta_{c}<\theta\leq \theta_{j}\\
0, &\theta > \theta_{j},
\end{cases}
\end{align}
where $\theta_{c}$ is introduced to avoid the divergence at $\theta=0$, and the power law index $k$ is taken to be the same for both the energy and Lorentz factor for simplicity. Geometric constraints impose the limit $\theta_{j} \leq \pi/2$, and $\theta_{c}$ is chosen to be much smaller than any of the other angles of interest, with a lower limit of $\theta_{c}>1/\Gamma_{\mathrm{max}}$. In theory, the universal structured jet model should restrict $k$ to $1.5\lesssim k \leq 2$ in order to recover the properties of the uniform jet model and to guarantee a standard energy reservoir~\citep{zhang_usj,rossi_usj}, but recent attempts to fit $k$ from data have given much larger values (e.g. $k\sim 8$, ~\cite{Pescalli}). For a given $\Gamma_{0}$ and $\theta_{j}$, $k$ is constrained so that $\Gamma(\theta_{j})\geq 1$:
\begin{align}
k_{\mathrm{max}} = \frac{\log{\Gamma_{0}}}{\log(\theta_{j}/\theta_{c})}.
\label{eq:kmax}
\end{align}

\begin{figure}
\centering
\centerline{\includegraphics[width=0.5\textwidth]{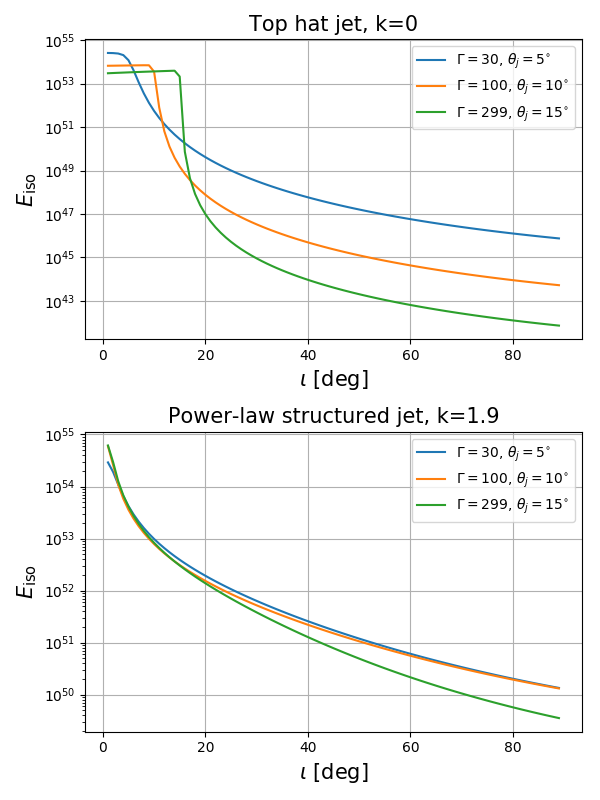}}
\caption{Isotropic equivalent energy as a function of inclination angle for the top-hat (top) and universal structured jet model with k=1.9 (bottom) for a range of opening angles and Lorentz factors for a jet with total energy $E_{0}^{\gamma} = 10^{52}~\mathrm{erg}$. }
\label{fig:Eiso_demonstration}
\end{figure}

The isotropic equivalent energy as a function of the inclination angle is shown in Fig~\ref{fig:Eiso_demonstration}. The energy drops off quickly once the inclination angle exceeds the opening angle of the jet, and higher Lorentz factors beam the emission more efficiently, making detection more difficult for off-axis observers. The shape of the profile is nearly indistinguishable at high inclination angles.

\section{Methods}
\label{sec:methods}
\subsection{Bayesian parameter estimation}
We aim to measure the posterior probability distributions for the opening angle, $\theta_{j}$, power law index, $k$, total source-frame gamma-ray energy $E_{0}^{\gamma}$, and Lorentz factor $\Gamma_{0}$ through the joint observation of electromagnetic and GW data. We define three relevant parameter sets:
\begin{itemize}
\item $\mathbf{x} = \{E_{0}^{\gamma}, \Gamma_{0}, \theta_{j},k\}$ -- Parameters unique to the electromagnetic data
\item $\boldsymbol{\eta} = \{\iota,d_{L},z\}$ -- Parameters common to both the electromagnetic and gravitational-wave data: orbital inclination, luminosity distance and redshift~\footnote{The gravitational-wave data only provides the luminosity distance, from which the redshift can be obtained if the cosmology is known.}
\item $\boldsymbol{\lambda}$ -- Parameters unique to the gravitational-wave data such as the masses and spins of the neutron stars
\end{itemize}

We then define two likelihood functions in terms of these parameters:
\begin{gather}
\mathcal{L}^\text{GW}(h|\boldsymbol{\eta},\boldsymbol{\lambda}) \propto \exp\left( -\sum_k \frac{ 2\left( h(f_k)-\widehat{h}(f_k|\boldsymbol{\eta},\boldsymbol{\lambda})\right)^2}{TS_h(f_k)^2}\right)\\
\mathcal{L}^\text{EM}(F^{\gamma}|\mathbf{x},\boldsymbol{\eta}) \propto \exp\left( - \frac{ \left( F^{\gamma} -\widehat{F^{\gamma}}(\mathbf{x},\boldsymbol{\eta})\right)^2}{2\sigma_{F^{\gamma}}^2}\right). \label{eq:em_likelihood}
\end{gather}
The first likelihood, ${\cal L}^\text{GW}(h|\boldsymbol{\eta},\boldsymbol{\lambda})$, is the gravitational-wave likelihood function for the strain data $h(f_k)$ in each frequency bin $f_{k}$ for an analysis segment of duration $T$ given the waveform model $\widehat{h}(f_k|\boldsymbol{\eta},\boldsymbol{\lambda})$~\citep{lalinference}. The second likelihood, $\mathcal{L}^\text{EM}(F^{\gamma}|\mathbf{x},\boldsymbol{\eta})$, is the electromagnetic likelihood function for the fluence data, which depends on the purely electromagnetic parameters and some of the binary parameters given by the subset $\boldsymbol{\eta}$. We assume that the fluence is measured with an uncertainty of $\sigma_{F^{\gamma}}=0.3\times10^{-7}\,[\mathrm{erg/cm^{2}}]$, which is the average reported fluence uncertainty for GRB 170817A~\citep{grb170817A_fermi, grb170817A_integral}. We assume the gravitational-wave noise is Gaussian, where $S_{h}(f_{k})$ is the noise power spectral density at Advanced LIGO design sensitivity~\citep{aLIGO}.

Combining the two likelihoods, we obtain a posterior for the EM parameters $\mathbf{x}$:
\begin{align}
p(\mathbf{x}\vert h,F^{\gamma}) &= \frac{\pi(\mathbf{x})}{\mathcal{Z}_{\mathbf{x}}}\int \Bigg[d\boldsymbol{\eta} \, \mathcal{L}^{\mathrm{EM}}(F^{\gamma}\vert \mathbf{x},\boldsymbol{\eta})\pi(\boldsymbol{\eta}) \label{eq:posterior}
\\ \nonumber
&\times \, \bigg(\int \mathcal{L}^{\mathrm{GW}}(h\vert \boldsymbol{\eta},\boldsymbol{\lambda})\pi(\boldsymbol{\lambda})d\boldsymbol{\lambda}\bigg)\Bigg]\\
&\equiv \frac{1}{\cal{Z}_{\mathbf{x}}} 
{\cal L}^\text{GW+EM}(h,F^{\gamma}|\mathbf{x})\pi(\mathbf{x})
\end{align}
where $\pi(\mathbf{x})$, $\pi(\boldsymbol{\eta})$, and $\pi(\boldsymbol{\lambda})$ are the priors for each set of parameters defined above.
In the first step we marginalize separately over the gravitational-wave nuisance parameters $\boldsymbol{\lambda}$~\footnote{We stress that we use the same symbol for the marginalized likelihood, just removing the marginalized parameters from the list of parameters that the likelihood depends on. So for example the marginalization over the gravitational-wave parameters $\boldsymbol{\lambda}$ follows from $\mathcal{L}^{\mathrm{GW}}(h | \boldsymbol{\eta})\equiv \int \mathcal{L}^{\mathrm{GW}}(h\vert \boldsymbol{\eta},\boldsymbol{\lambda})\pi(\boldsymbol{\lambda})d\boldsymbol{\lambda}$.} and the common parameters $\boldsymbol{\eta}$, and in the second step we define the joint GW+EM likelihood:
\begin{align}
    \mathcal{L}^{\mathrm{GW+EM}}(h,F^{\gamma}|\mathbf{x}) = \int \mathcal{L}^{\mathrm{GW}}(h | \boldsymbol{\eta}) \mathcal{L}^{\mathrm{EM}}(F^{\gamma} | \mathbf{x},\boldsymbol{\eta})\pi(\boldsymbol{\eta}) d\boldsymbol{\eta}.
    \label{eq:joint_likelihood}
\end{align}
${\cal Z}_{\mathbf{x}}$ is the Bayesian evidence obtained by marginalizing the joint likelihood over the GRB parameters $\mathbf{x}$:
\begin{align}
    {\cal Z}_{\mathbf{x}} &= \int {\cal L}^\text{GW+EM}(h,F^{\gamma}|\mathbf{x})\pi(\mathbf{x})d\mathbf{x}.
    \label{eq:em_evidence}
\end{align}

\begin{comment}
If we assume that the parameters $\mathbf{x} = \{E_{0}^{\gamma}, \Gamma_{0}, \theta_{j},k\}$ are universal properties of all gamma-ray bursts, then we may construct a joint posterior using the product of the joint likelihoods from $N$ events:
\begin{align}
p_\text{tot}(\mathbf{x}|\{h\}, \{F^{\gamma}\}) = \frac{\pi(\mathbf{x})}{{\cal Z}_N} \prod_{j=1}^N 
\mathcal{L}^\text{GW+EM}(h_j,F^{\gamma}_j|\mathbf{x})
\label{eq:stacked_post}
\end{align}
\end{comment}

\subsection{Simulated coincident event population}
We simulate 100 binary neutron star gravitational-wave events. The masses are drawn uniformly in chirp mass:
\begin{align}
\mathcal{M} = \frac{(m_{1}m_{2})^{3/5}}{(m_{1}+m_{2})^{1/5}}, 
\end{align}
between 0.888 and $1.63~\Msun$ and in mass ratio: 
\begin{align}
q = m_{2}/m_{1},\ m_{1}\geq m_{2}, 
\end{align}
between 0.7 and 1.
These ranges are chosen to be consistent with the domain of validity of the reduced order quadrature model~\citep{roq} for the IMRPhenomPv2~\citep{imrphenomp} waveform, which we employ to keep the computational cost under control.

The events are distributed uniformly in comoving volume between 10 and 80~Mpc and added into a Hanford-Livingston detector network using simulated design sensitivity Gaussian noise. While this network will be sensitive to BNS mergers at larger distances out to $\sim 200~\mathrm{Mpc}$~\citep{observing_scenarios}, we choose to limit the maximum event distance for this analysis so that all the simulated events are detectable.
We comment on the consequences of this setup in Sec.~\ref{sec:discussion}. At these small distances, the redshift is calculated from the luminosity distance posterior as $z = d_{L}/D_{H}$, where $D_{H} = 9.26\times 10^{27}/h_{0}~\mathrm{cm}$ is the Hubble distance, and we take $h_{0} = 0.68$. The inclination angle and sky position are distributed isotropically. The priors used when running the sampler, $\pi(\boldsymbol{\lambda})$ and $\pi(\boldsymbol{\eta})$, are identical to those from which the event parameters are drawn. The neutron stars are assumed to be non-spinning point masses with no tidal deformability, which does not have a significant effect on the inference of the common GW+EM parameters, $\boldsymbol{\eta}$. 

\begin{comment}
\begin{figure}
\centering
\centerline{\includegraphics[width=0.5\textwidth]{gw_snr_distribution_paper_wide.png}}
\caption{Distribution of optimal network SNR for the 100 simulated binary neutron star gravitational-wave events included in this analysis, injected into a Hanford-Livingston network at design sensitivity. 
%
}
\label{fig:gw_snr}
\end{figure}
\end{comment}

Each of the 100 simulated gravitational-wave sources is randomly associated with a GRB event. We simulate two GRB populations, each with 100 events -- one with only top-hat jets and one with power-law jets with $k=1.9$. These fiducial models are chosen to demonstrate our method, and we leave consideration of other structures like the Gaussian jet to future work. In both cases, the energy is drawn from a truncated log-normal distribution in $\log_{10}{E_{0}}$ between $10^{47}$ and $10^{54}~\mathrm{erg}$ with a mean of $10^{50}~\mathrm{erg}$ and a width of one dex. For the top-hat population, the opening angle $\theta_{j}$ is drawn from a truncated Gaussian between $2^{\circ}$ and $50^{\circ}$ with a mean of $25^{\circ}$ and a width of $5^{\circ}$, and the Lorentz factor is also drawn from a truncated Gaussian with $\mu_{\Gamma} = 100,\ \sigma_{\Gamma} = 50,\ 2 \leq \Gamma \leq 299$. For the power-law population, the distributions of $\Gamma$ and $\theta_{j}$ are chosen to preserve the constraint imposed by Eq.~\ref{eq:kmax} for $k=1.9$. Both the Lorentz factor and opening angle are drawn from truncated Gaussians with the same boundaries described above and $\mu_{\Gamma} = 270,\ \sigma_{\Gamma} = 20,\ \mu_{\theta_{j}} = 7^{\circ},\ \sigma_{\theta_{j}} = 4^{\circ}$. The distributions used to simulate the top-hat and power-law populations are summarized in Tables~\ref{tab:top-hat} and \ref{tab:power-law}, respectively. In all cases the parameter boundaries are chosen to be consistent with theory and the observed population of short GRBs.
\begin{table}
    \centering
\begin{tabular}{|p{1cm} ||p{1cm} p{1cm} p{1cm} p{1cm}|}
    \hline
     & $\mu$ & $\sigma$ & min & max \\
    \hline \hline
    $\log{E_{0}}$ & 50 & 1 & 47 & 54 \\
    $\Gamma$ & 100 & 50 & 2 & 299 \\
    $\theta_{j}$ & $25^{\circ}$ & $5^{\circ}$ & $2^{\circ}$ & $50^{\circ}$ \\
    $\mu_{k}$ & 0 & 0 & 0 & 0 \\
    \hline
\end{tabular}
\caption{Parameters describing the distributions used for simulating the population of top-hat jets.}
\label{tab:top-hat}
\begin{tabular}{|p{1cm} ||p{1cm} p{1cm} p{1cm} p{1cm}|}
    \hline
     & $\mu$ & $\sigma$ & min & max \\
    \hline \hline
    $\log{E_{0}}$ & 50 & 1 & 47 & 54 \\
    $\Gamma$ & 270 & 20 & 2 & 299 \\
    $\theta_{j}$ & $7^{\circ}$ & $4^{\circ}$ & $2^{\circ}$ & $50^{\circ}$ \\
    $\mu_{k}$ & 1.9 & 0 & 0 & 0 \\
    \hline
\end{tabular}
\caption{Parameters describing the distributions used for simulating the population of power-law jets.}
\label{tab:power-law}
\end{table}

The fluence for each joint GW+EM event is calculated by evaluating the expression in Eq.~\ref{eq:fluence} at the simulated event parameters. Since the integral is costly to evaluate analytically, we use a lookup table to calculate it efficiently, see Appendix~\ref{sec:AppendixIntegral}.

\begin{figure}
\centering
\centerline{\includegraphics[width=0.5\textwidth]{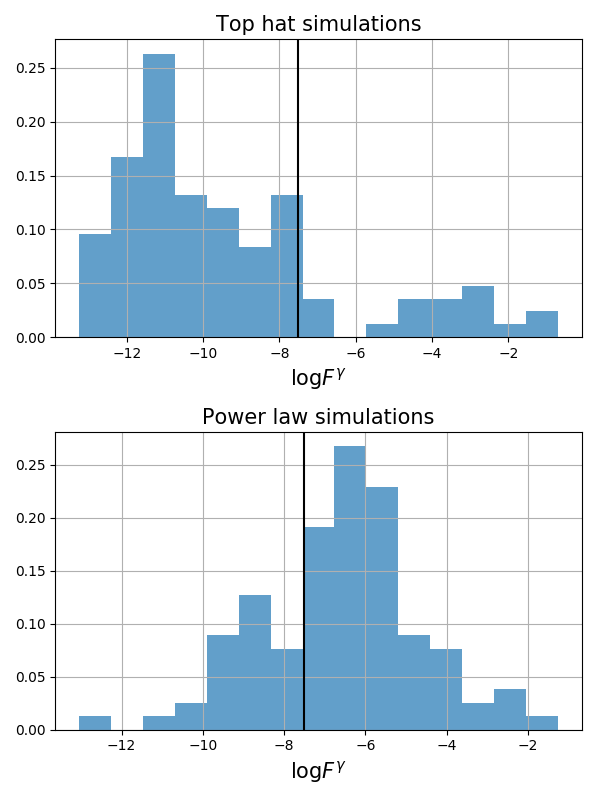}}
\caption{Distributions of $\log{F^{\gamma}}$ for the top-hat (left) and power-law (right) simulated populations. The vertical line indicates the value of $\sigma_{F^{\gamma}}$, which serves as a threshold for the detection of the sGRB emission. There are 21 detectable top-hat events and 73 detectable power-law events out of the 100 events in each population.}
\label{fig:true_fluence}
\end{figure}

To simulate the GRB detector noise, the ``measured" value of the fluence for each GW+EM event is drawn from a Gaussian distribution centered on the ``true" fluence value calculated as described above for each event's simulated parameters with the width given by $\sigma_{F^{\gamma}}$. The distribution of ``true" fluences is shown in Fig.~\ref{fig:true_fluence}. This means that some events with sub-threshold ``true" fluence values will end up with negative values for the ``measured" fluence, which corresponds to a dearth of counts after background subtraction. While the gamma-ray photons arriving at the GRB detector are actually Poisson-distributed, the Gaussian approximation we make in simulating the detector noise and in the likelihood in Eq.~\ref{eq:em_likelihood} is valid in the limit of large numbers of counts. Because we run our analysis only on detectable BNS gravitational-wave events but include non-detections of associated short GRBs when the ``measured" fluence is sub-threshold, this corresponds to a GW-triggered search including upper limits on fluence obtained by GRB satellites. This does not include GW events for which the sky region is outside the GRB satellite's field of view or cases where the GRB satellites are not in observing mode at the time of the GW trigger. 

For each event, we obtain posteriors for the GW parameters $(\boldsymbol{\eta},\boldsymbol{\lambda})$ and the EM-only parameters  $\mathbf{x}$ using the \texttt{bilby} parameter estimation package~\citep{bilby} and the \texttt{dynesty} nested sampler~\citep{dynesty}. Because we marginalize over the uncertainty in the GW parameters following the prescription detailed in Appendix~\ref{sec:AppendixBayesian}, the posteriors for the EM-only parameters include the effects of the correlation between distance and inclination in the gravitational-wave posteriors and the uncertainty in these parameters.  We use uniform priors, $\pi(\mathbf{x})$, for all parameters in the ranges covered by the simulated event distributions (see Tables~\ref{tab:top-hat} and \ref{tab:power-law}) except for $k$, which has a conditional prior that is uniform between 0 and $k_{\mathrm{max}}$ as defined in Eq.~\ref{eq:kmax} for each prior sample in $\Gamma$ and $\theta_{j}$.

\subsection{Hierarchical modeling}
While each GRB will have a different value of the energy, Lorentz factor, and opening angle, we can use the population of events to measure the properties of the underlying distributions of which the individual parameters are a representative sample. This is  usually referred to as hierarchical modeling. We assume that the underlying distribution for the parameters $\mathbf{x}$ can be characterized by a set of hyper-parameters $\boldsymbol{\Lambda} = \{ \mu_{E_{0}^{\gamma}}, \sigma_{E_{0}^{\gamma}}, \mu_{\Gamma_{0}}, \sigma_{\Gamma_{0}}, \mu_{\theta_{j}}, \sigma_{\theta_{j}}, \mu_{k}, \sigma_{k}\}$, i.e. we assume individual GRB sources have parameters drawn from truncated Gaussian distributions with unknown means and standard deviations. This underlying distribution is called the hyper-prior, $\pi(\mathbf{x} | \boldsymbol{\Lambda})$.
We stress that while we have fixed the true value of $k$ to be the same for all simulated GRB events ($k=0$ for the top-hat model and $k=1.9$ for the power-law model) we still measure the hyper-parameters associated with $k$ in order to determine whether a universal angular emission profile can be inferred from the data. For individual events, the joint likelihood in Eq.~\ref{eq:joint_likelihood} depends on the hyper-parameters only implicitly through the distributions of the individual-event parameters $\mathbf{x}$. The likelihood for the hyper-parameters is thus obtained by marginalizing the joint likelihood for the $j$th event over the EM-only parameters:
\begin{align}
\mathcal{L}(h_{j}, F^{\gamma}_{j} \vert \boldsymbol{\Lambda}) &= \int d\mathbf{x}\  \mathcal{L}^{\mathrm{GW+EM}}(h_{j}, F^{\gamma}_{j} \vert \mathbf{x}, \boldsymbol{\Lambda})\pi(\mathbf{x} \vert \boldsymbol{\Lambda}).
\label{eq:hyper_likelihood}
\end{align}

Hierarchical modeling must typically take selection biases into account due to the fact that only detected events, which have different properties than the population as a whole, are included in the analysis sample, thus affecting the inference of the hyper-parameters~\citep{Loredo_2004, Abbott2016, Fishbach_2018, Wysocki_2019, Mortlock_2018, Mandel_2019}. However, we do not introduce a cut based on the detectability of either the gravitational-wave or electromagnetic data so that we don't need to account for selection biases in our analysis. Even if a cut were introduced on the gravitational-wave data, this would not affect the inference of the hyper-parameters $\boldsymbol{\Lambda}$ describing the electromagnetic parameters $\mathbf{x}$, as explained in detail in Appendix~\ref{sec:AppendixBayesian}. Therefore, the posterior on the hyper-parameters for an ensemble of $N$ events is obtained by multiplying the individual event likelihoods without any modifications to account for the probability of detection:
\begin{align}
p(\boldsymbol{\Lambda} \vert \{h\}, \{F^{\gamma}\})&= \frac{\pi(\boldsymbol{\Lambda})}{\mathcal{Z}_{\boldsymbol{\Lambda}}}\prod_{j}^{N}\mathcal{L}(h_{j}, F^{\gamma}_{j} \vert \boldsymbol{\Lambda}),
\label{eq:hyper_posterior}
\end{align}
where $\pi(\boldsymbol{\Lambda})$ is the prior on the hyper-parameters given in Table~\ref{tab:hyper_priors}, $\mathcal{L}(h_{j}, F^{\gamma}_{j} \vert \boldsymbol{\Lambda})$ is the joint EM-GW likelihood marginalized over the hyper-prior for the $j$-th event from Eq.~\ref{eq:hyper_likelihood}, and $\mathcal{Z}_{\boldsymbol{\Lambda}}$ is the hyper-evidence (see Appendix~\ref{sec:AppendixBayesian}).
We produce samples from this distribution using the \texttt{bilby} implementation of the \texttt{pymultinest}~\citep{Feroz_2008, Feroz_2009, Feroz_2013, Buchner_2014} and \texttt{cpnest} samplers~\citep{cpnest}. 
\begin{table}
    \centering
\begin{tabular}{|p{1cm} ||p{2cm} p{1cm} p{1cm}|}
    \hline
     & Shape & min & max \\
    \hline \hline
    $\mu_{\log{E_{0}}}$ & Uniform & 47 & 54 \\
    $\sigma_{\log{E_{0}}}$ & Uniform & 0.1 & 5 \\
    $\mu_{\Gamma}$ & Uniform & 2 & 299 \\
    $\sigma_{\Gamma}$ & Uniform & 5 & 200 \\
    $\mu_{\theta_{j}}$ & Uniform & $2^{\circ}$ & $50^{\circ}$ \\
    $\sigma_{\theta_{j}}$ & Uniform & $1^{\circ}$ & $15^{\circ}$ \\
    $\mu_{k}$ & Uniform & 0 & 8 \\
    $\sigma_{k}$ & Log-Uniform & $10^{-4}$ & 1 \\
    \hline
\end{tabular}
\caption{Priors on the hyper-parameters used in the hierarchical modeling step, $\pi(\boldsymbol{\Lambda})$.}
\label{tab:hyper_priors}
\end{table}

\section{Results}
\label{sec:results}
\subsection{Individual event analysis}
\label{sec:individual}
\begin{figure}
\centering
\centerline{\includegraphics[width=0.5\textwidth]{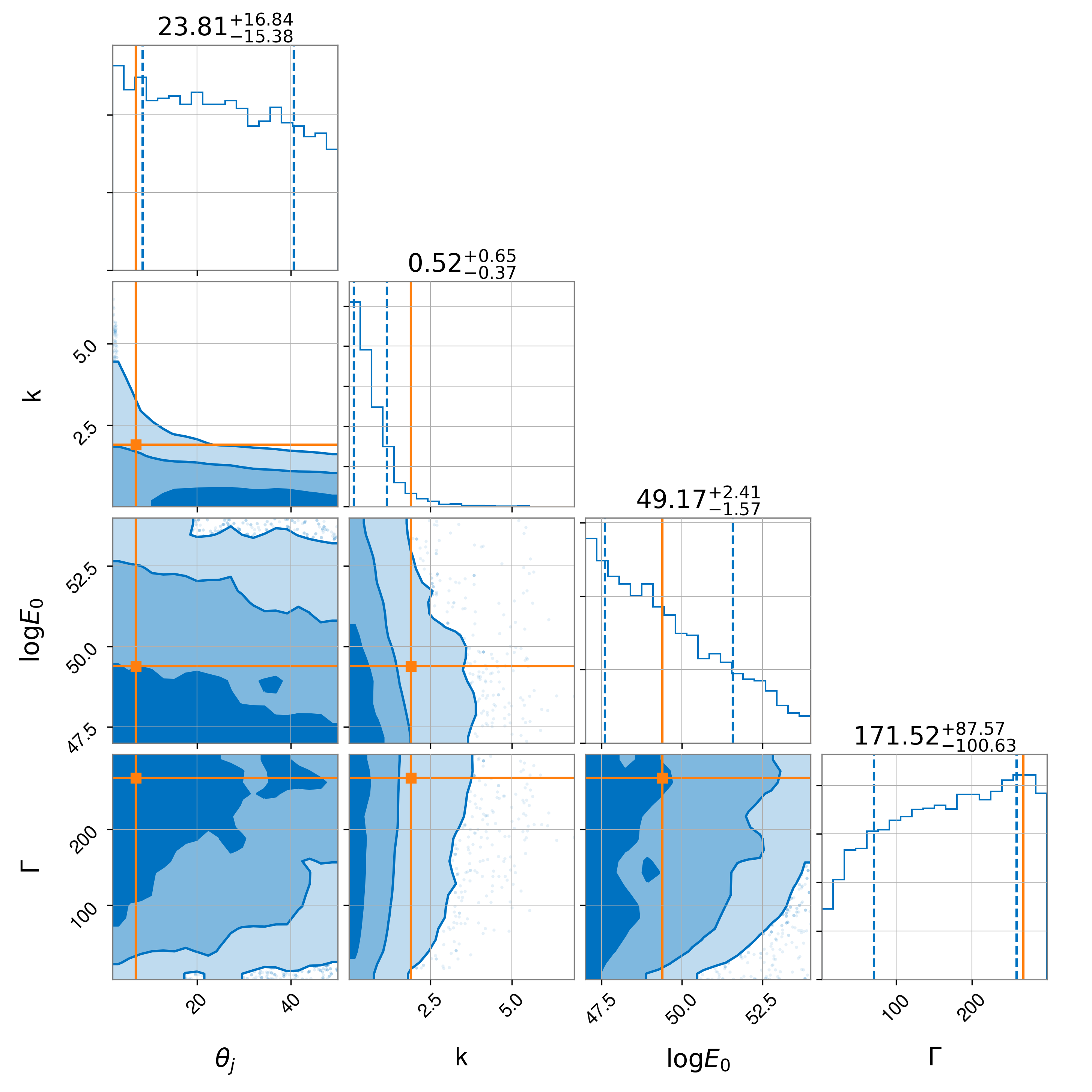}}
\caption{Corner plot for an uninformative power-law event with true fluence $F^{\gamma} = 2.88\times 10^{-9}~\mathrm{erg/cm^{2}}$. The orange lines represent the true parameter values, while the blue dashed lines are the $1\sigma$ uncertainties. These are also indicated above each marginalized posterior along with the median value for each parameter.}
\label{fig:uninformative_power_law}
\end{figure}

The morphology of the individual event posteriors for the EM-only parameters $\mathbf{x}$ varies depending on the SNR of the GRB signal. The corner plot for an uninformative (sub-threshold) power-law event is shown in Fig.~\ref{fig:uninformative_power_law}. The posterior for $k$, which is highly peaked around $k=0$, essentially returns the prior for this uninformative event. Additionally, there is more support for higher values of $k$ for narrower opening angles and higher Lorentz factors, since the $k_{\mathrm{max}}$ condition is more easily satisfied in that part of the parameter space. The energy posterior favors lower values since a lower energy results in a lower fluence, and the posterior for the Lorentz factor slightly favors higher values because this causes the fluence to drop off more steeply for inclination angles outside the jet edge (see Fig.~\ref{fig:Eiso_demonstration}). The posteriors for uninformative top-hat events show similar trends.
\begin{comment}
\begin{figure}
\centering
\centerline{\includegraphics[width=0.5\textwidth]{sgrb_realistic_23_uniformAll_corner.png}}
\caption{Corner plot for an uninformative top-hat event with true fluence $F^{\gamma} = 4.81\times 10^{-10}~\mathrm{erg/cm^{2}}$. The orange lines represent the true parameter values, while the blue dashed lines are the $1\sigma$ uncertainties. These are also indicated above each marginalized posterior along with the median value for each parameter.}
\label{fig:uninformative_top_hat}
\end{figure}
\end{comment}

\begin{figure}
\centering
\centerline{\includegraphics[width=0.5\textwidth]{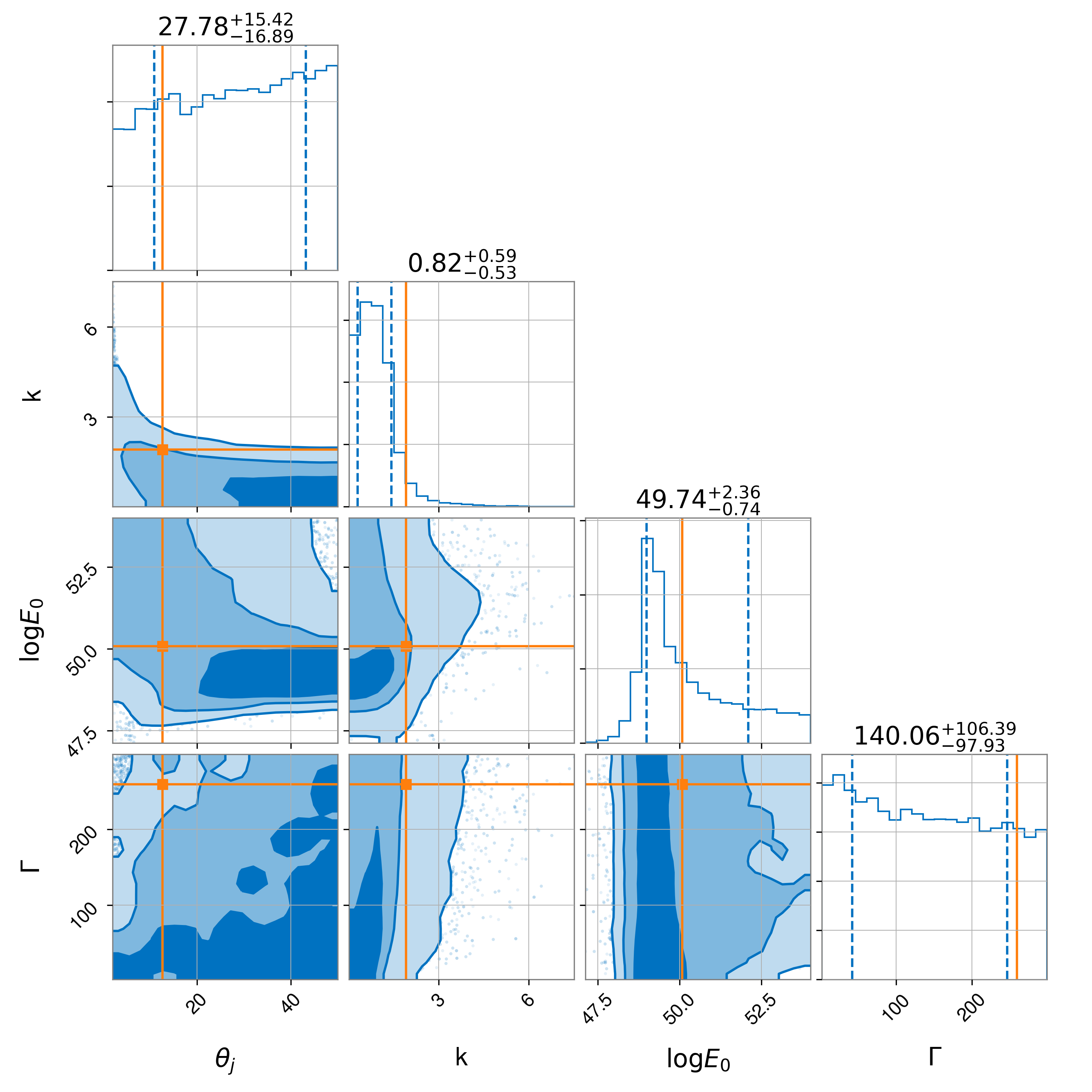}}
\caption{Corner plot for an informative power-law event with true fluence $F^{\gamma} = 1.46\times 10^{-4}~\mathrm{erg/cm^{2}}$.}
\label{fig:informative_power_law}
\end{figure}

For informative events, the best-constrained parameter is $E^{\gamma}_{0}$, since it decouples from the integral expression encoding the dependence of the fluence on the other three parameters in Eq.~\ref{eq:fluence}. The corner plot for a relatively informative power-law event is shown in Fig.~\ref{fig:informative_power_law}. Because we are trying to constrain four parameters with only one piece of data (the fluence), there are degenerate regions of parameter space that can produce the same fluence value. A wider opening angle but a steeper drop-off of the fluence due to a higher Lorentz factor could yield the same fluence value as a narrower opening angle with a more gradual drop-off for a particular inclination. A higher value of $k$ also causes the fluence to drop off more quickly, which could be compensated for by increasing the energy of the event. Because of these parameter-space degeneracies, we only observe very weak deviations from the prior in the posteriors for $\Gamma$ and $\theta_{j}$. The posteriors for informative top-hat events look very similar to the power-law posteriors, since the prior for $k$ very strongly disfavors values of $k\gtrsim 2$, so we do not show an example corner plot here.
\begin{comment}
\begin{figure}
\centering
\centerline{\includegraphics[width=0.5\textwidth]{sgrb_realistic_21_uniformAll_corner.png}}
\caption{Corner plot for a relatively informative top-hat event with true fluence $F^{\gamma} = 1.34\times 10^{-3}~\mathrm{erg/cm^{2}}$}
\label{fig:informative_top_hat}
\end{figure}
\end{comment}

\subsection{Population analysis}
While individual event posteriors are not very informative for the parameters encoding the jet geometry even for events with a high GRB SNR, we can use a population of events to place constraints on the hyper-parameters.

\subsubsection{Simulated top-hat population}
\begin{figure}
\centering
\centerline{\includegraphics[width=0.5\textwidth]{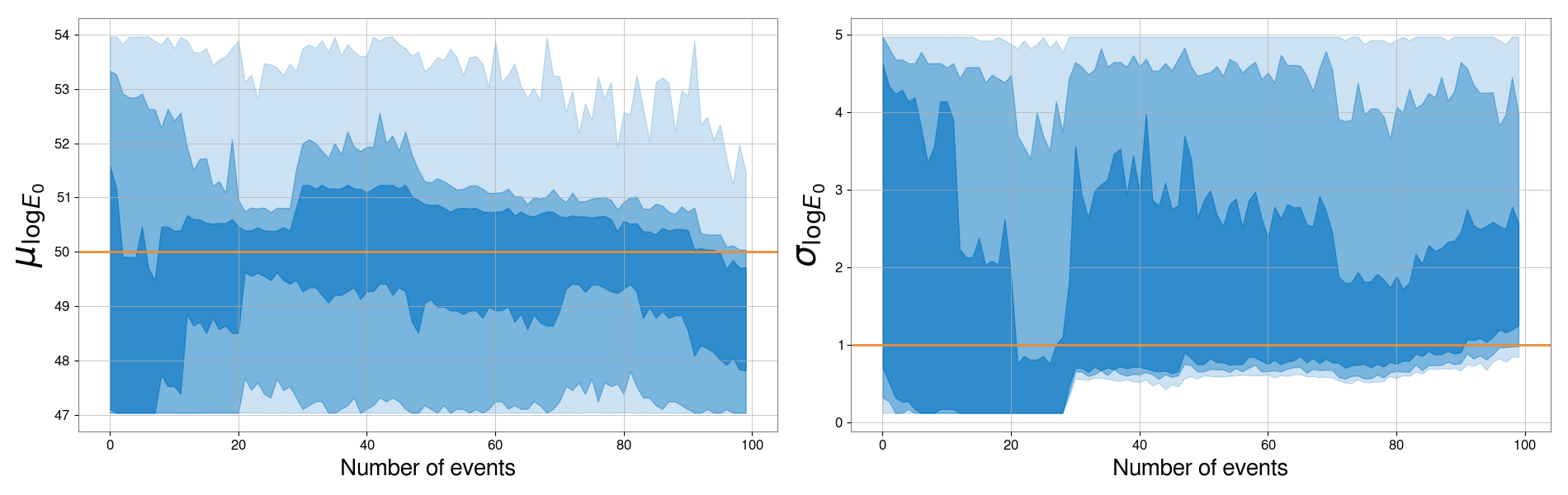}}
\centerline{\includegraphics[width=0.5\textwidth]{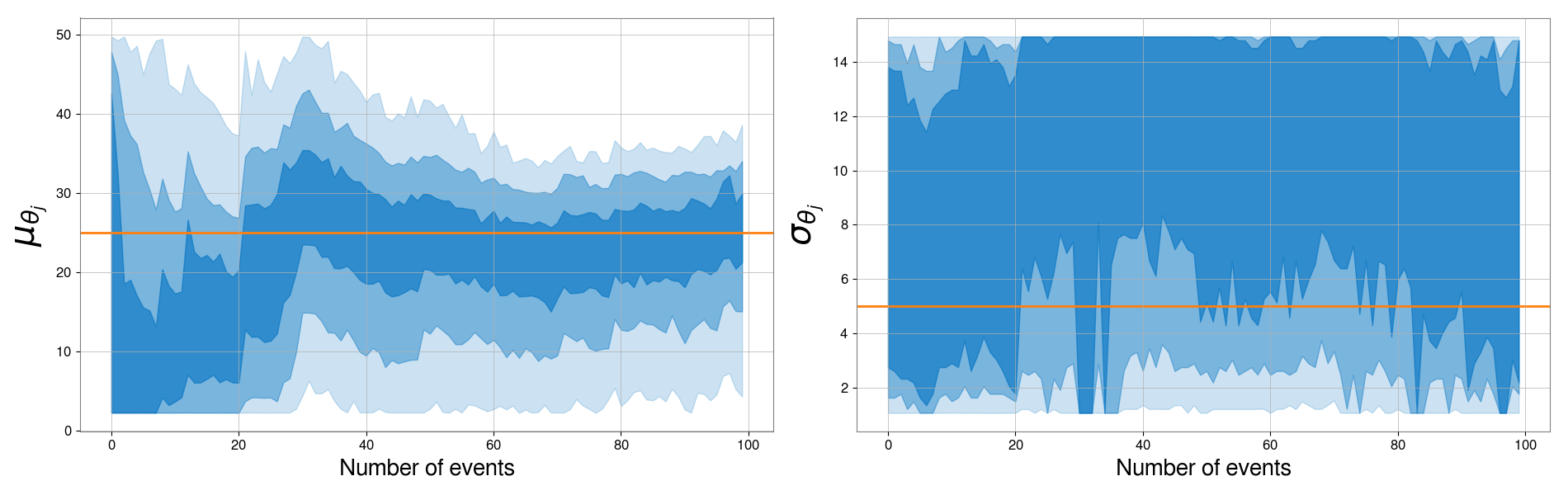}}
\centerline{\includegraphics[width=0.5\textwidth]{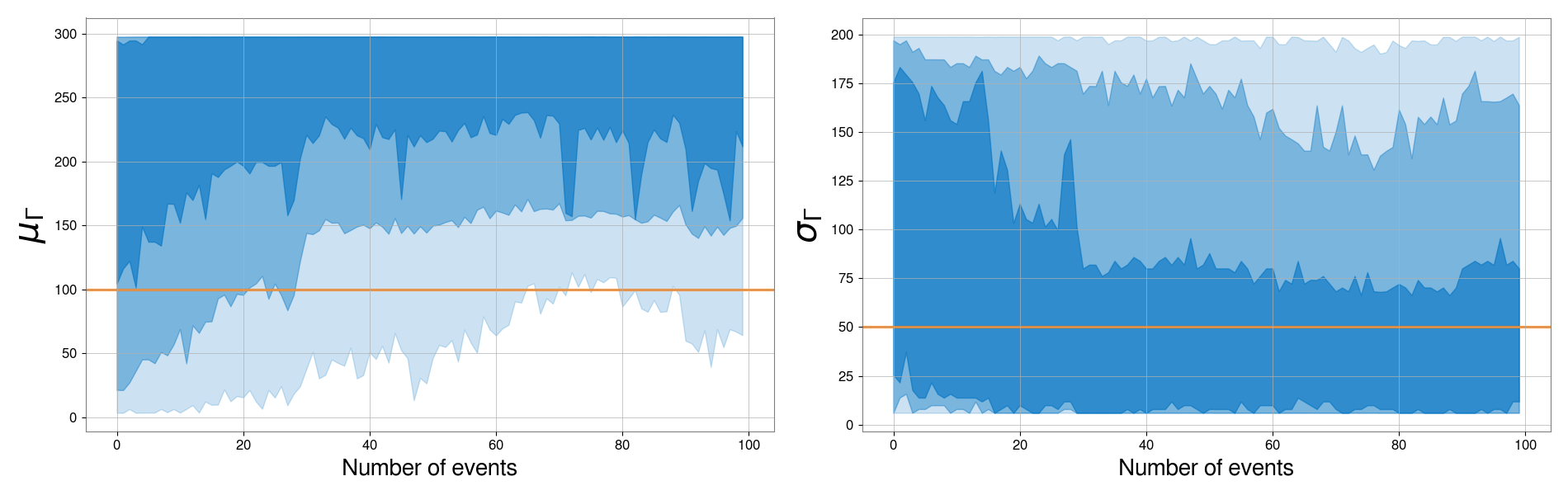}}
\centerline{\includegraphics[width=0.5\textwidth]{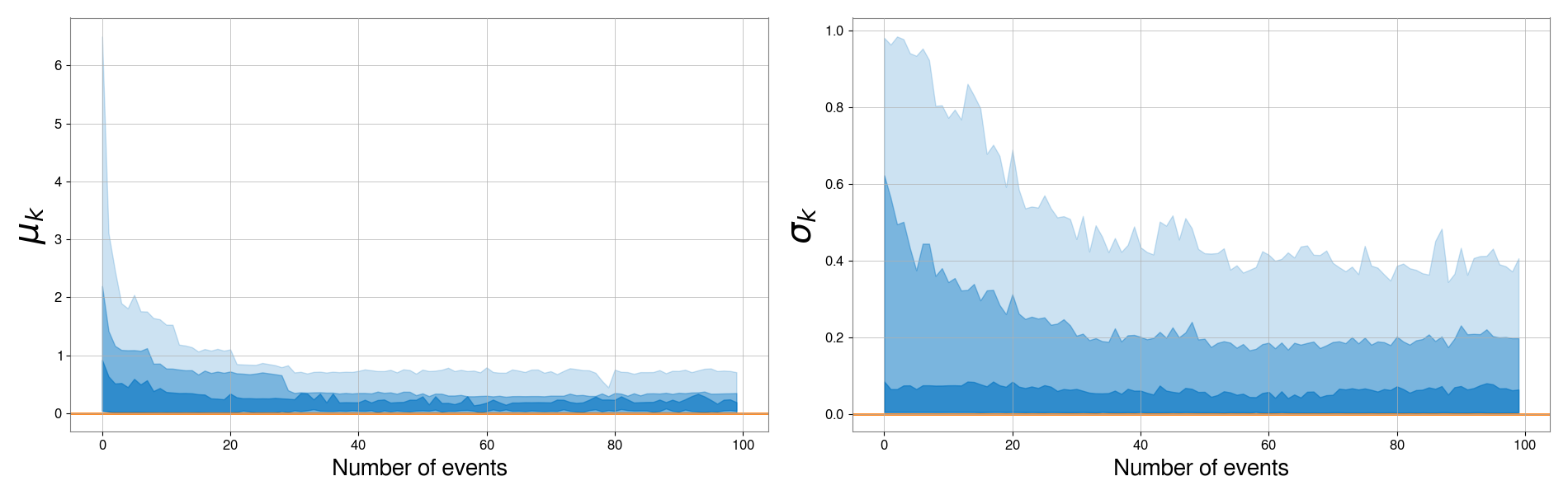}}
\caption{1, 2, and $3\sigma$ intervals for all hyper-parameters for the top-hat population, with the true value shown in orange.}
\label{fig:samplek_hyper_parameters}
\end{figure}

Fig.~\ref{fig:samplek_hyper_parameters} shows the 1, 2, and $3\sigma$ confidence intervals for the 8 hyper-parameters in our model for the top-hat population as a function of the number of BNS events accompanied by a GRB fluence measurement or upper limit included in the analysis. The energy hyper-parameters are well constrained to within $\sim 2~\mathrm{dex}$ in $\log{E_{0}}$ at the $1\sigma$ level with relatively few events, which is consistent with the energy being the most informative parameter in the individual event analysis presented earlier. The true values of both $\mu_{\log{E_{0}}}$ and $\sigma_{\log{E_{0}}}$ are contained within the $2\sigma$ confidence interval for all 100 events in the population. 
The hyper-parameter $\mu_{\theta_{j}}$ is also well constrained, with the $1\sigma$ region spanning about $10^{\circ}$, even though relatively little information can be gained about the opening angle of individual GRBs from the first step of PE.

The posterior for the $\sigma_{\theta_{j}}$ parameter is less informative, due in part to the fact that the prior range is narrower. The $\mu_{\Gamma}$ posterior is slightly offset towards higher values of $\Gamma$ because of the shape of the $\Gamma$ posterior for the uninformative individual events, which dominate the population. As described in the previous section, higher Lorentz factors lead to lower fluence values for observers outside the jet edge. The true value of $\mu_{\Gamma}$ is contained within the $3\sigma$ confidence interval, however. We note that the posterior for $\mu_{\Gamma}$ is also strongly dependent on the particular realization of the hyper-prior for the 100 true values that were chosen for our simulated population. We repeated the analysis with 100 different draws from the hyper-prior, and the offset in this parameter disappeared. This adds support to the idea that this posterior could converge to the true value with a larger population of events. The posterior for $\sigma_{\Gamma}$ is more informative because deviations from the uninformative posterior in Fig. ~\ref{fig:uninformative_power_law} indicate the spread of the true $\Gamma$ values. 
The true values of both $\mu_{k}$ and $\sigma_{k}$ are included in the $1\sigma$ confidence intervals for these parameters. 

\begin{figure}
\centering
\centerline{\includegraphics[width=0.5\textwidth]{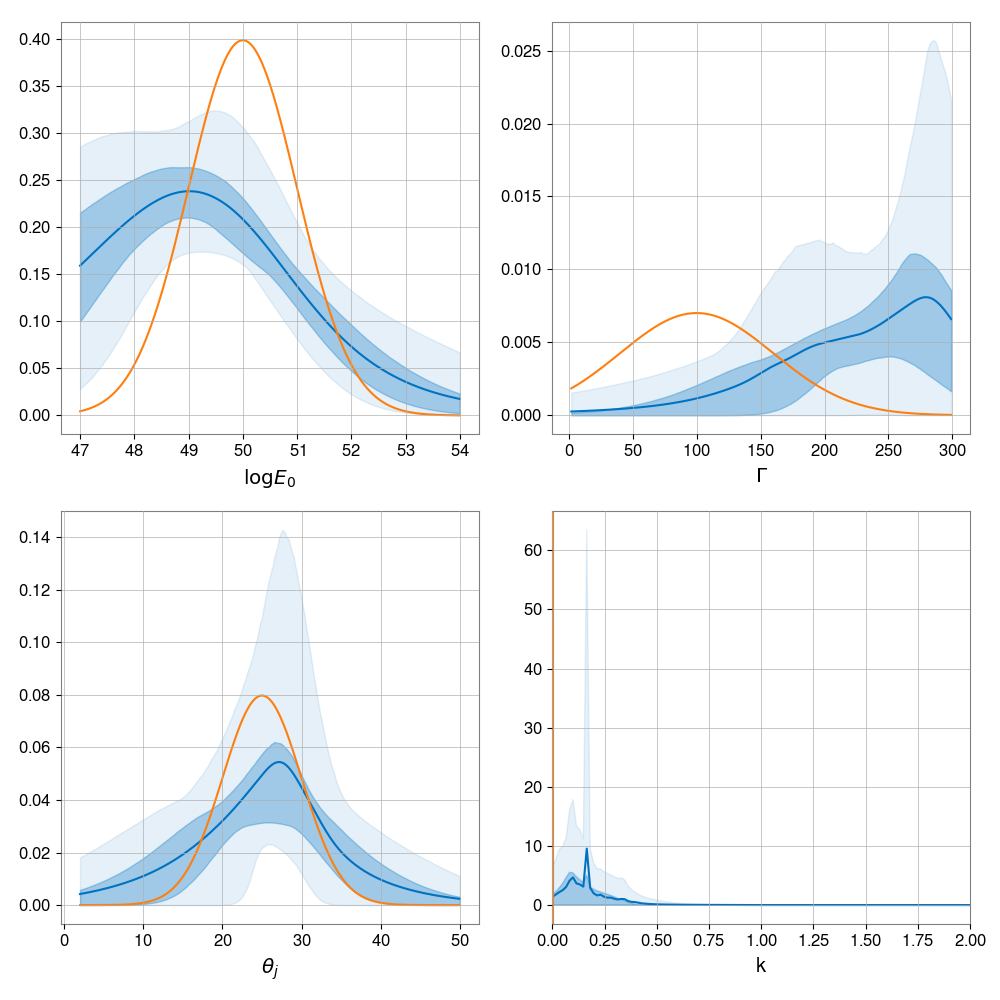}}
\caption{Posterior predictive distributions for the EM-only parameters for the top-hat population (dark blue lines) along with the 50\% and 90\% confidence intervals and the true distributions (orange lines).}
\label{fig:ppd_samplek}
\end{figure}

We also calculate the posterior predictive distributions (PPDs) for the parameters $\mathbf{x}$ from their hyper-parameters,
\begin{align}
    p_{\boldsymbol{\Lambda}}(\mathbf{x}|\{d\}, \{F^{\gamma}\}) = \int p(\boldsymbol{\Lambda} | \{d\}, \{F^{\gamma}\}) \pi(\mathbf{x} | \boldsymbol{\Lambda})d\boldsymbol{\Lambda},
\end{align}
which can be written for a discrete set of $m$ hyper-parameter posterior samples as:
\begin{align}
    p_{\boldsymbol{\Lambda}}(\mathbf{x}|\{d\}, \{F^{\gamma}\}) = \frac{1}{m}\sum_{i}^{m} \pi(\mathbf{x} | \boldsymbol{\Lambda}_{i}).
\end{align}
The posterior predictive distribution represents the updated prior on $\mathbf{x}$ after incorporating the information gained from the data via the posteriors on the hyper-parameters $\boldsymbol{\Lambda}$~\citep{bbh_pop}. The PPDs for the $\mathbf{x}$ parameters are shown in Fig.~\ref{fig:ppd_samplek} using the hyper-parameter posteriors inferred from all 100 events in our simulated population along with the 50\% and 90\% confidence regions and the true distributions used for simulating the events. As expected from the hyper-parameter posteriors presented in Fig.~\ref{fig:samplek_hyper_parameters}, the distribution for $\theta_{j}$ is the best recovered, and the distribution for $\Gamma$ peaks above the true value. The distribution for $k$ peaks away from 0 but is consistent with the true value within error, and the width of the PPD can be attributed to the sampling error encompassed in non-zero values of $\sigma_{k}$. The PPD for $\log{E_{0}}$ is slightly wider with a lower peak than the true value, consistent with the hyper-parameter posteriors for $\mu_{\log{E_{0}}}$ and $\sigma_{\log{E_{0}}}$.

\subsubsection{Simulated power-law population}
\begin{figure}
\centering
\centerline{\includegraphics[width=0.5\textwidth]{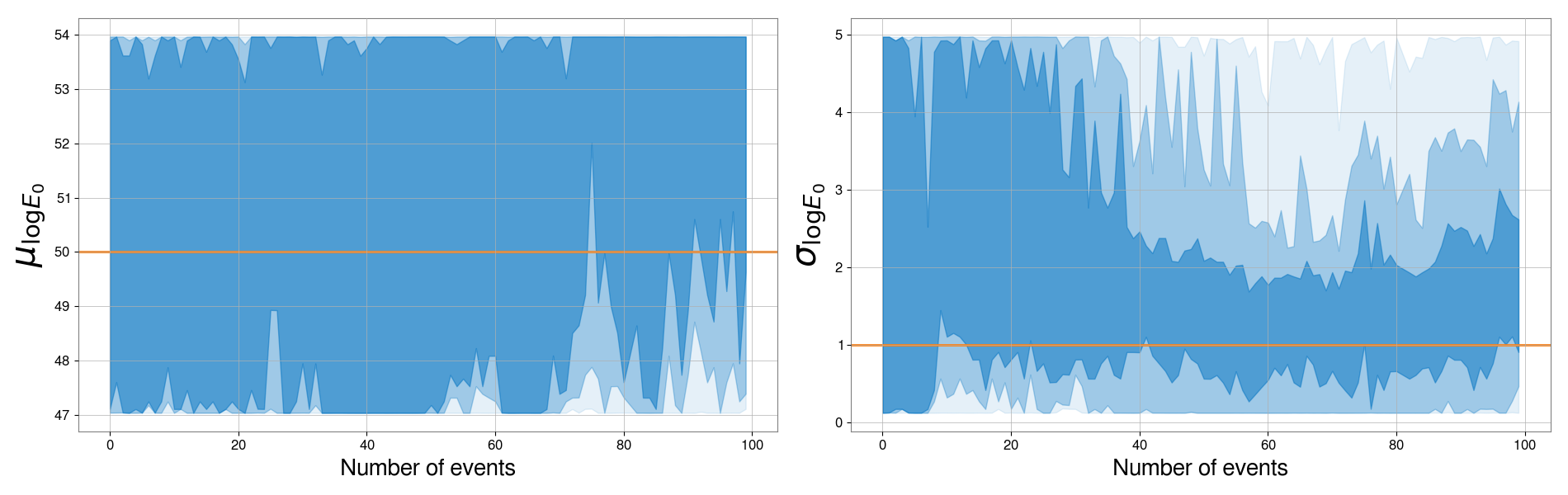}}
\centerline{\includegraphics[width=0.5\textwidth]{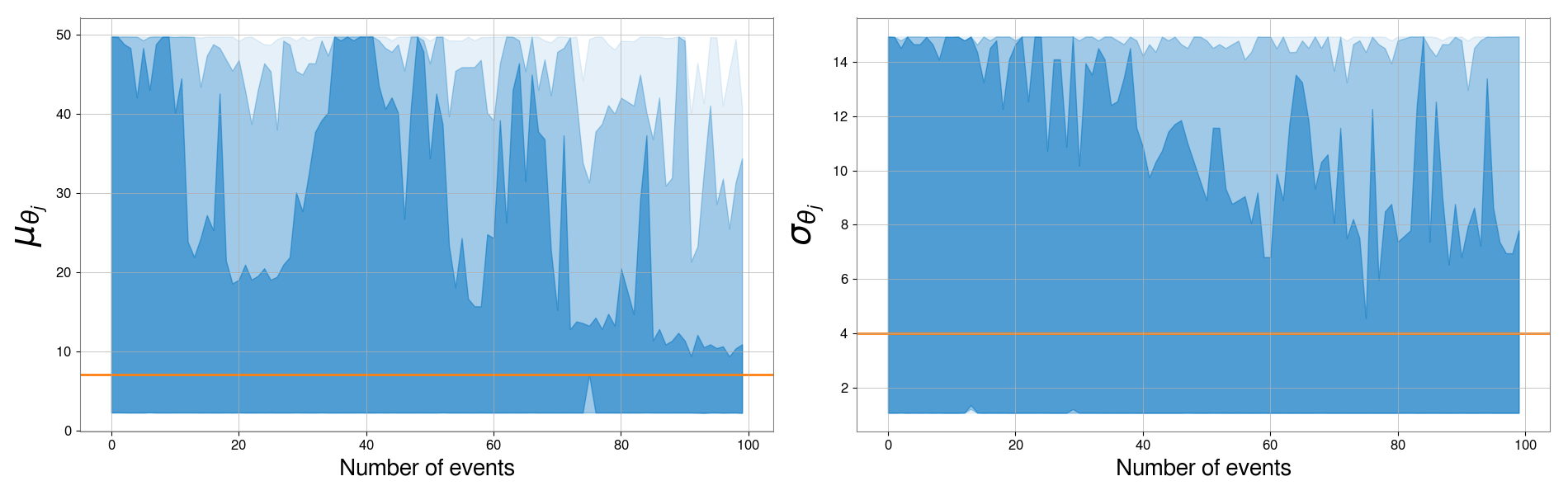}}
\centerline{\includegraphics[width=0.5\textwidth]{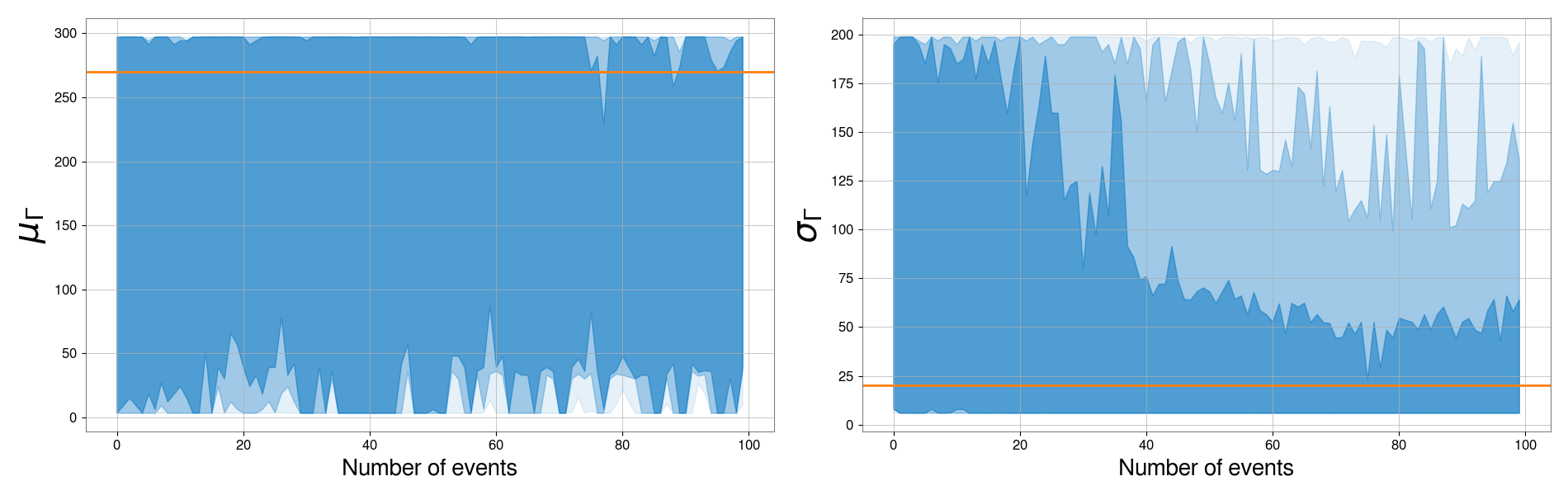}}
\centerline{\includegraphics[width=0.5\textwidth]{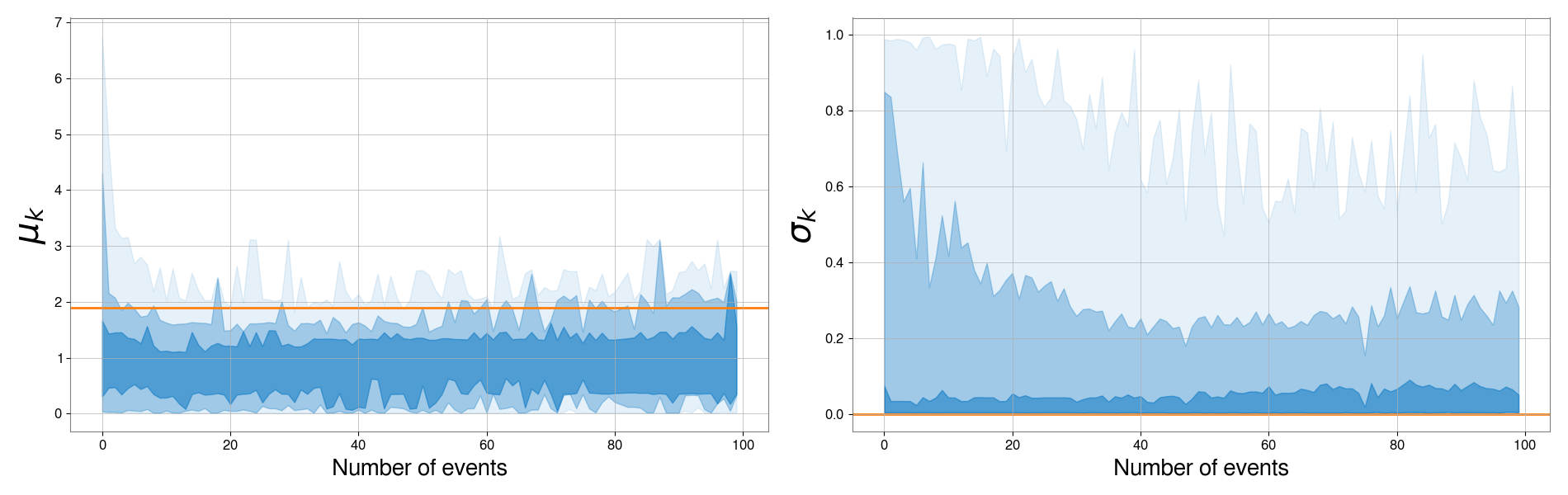}}
\caption{1, 2, and $3\sigma$ intervals for all hyper-parameters for the power-law population, with the true value shown in orange.}
\label{fig:power_law_hyper_parameters}
\end{figure}

Fig.~\ref{fig:power_law_hyper_parameters} shows the 1, 2, and $3\sigma$ regions for all 8 hyper-parameters for the power-law population. The mean of the energy distribution is wider than for the top-hat simulations, while the width is constrained at a similar level. 
The $1\sigma$ region for the $\mu_{\theta_{j}}$ posterior is constrained to $< 10^{\circ}$ and includes the true value, and the $\sigma_{\theta_{j}}$ posterior is again less informative. 
$\sigma_{\Gamma}$ is constrained to within $\sim 70$ at the $1\sigma$ level, which is slightly narrower than in the top-hat case even though the $\mu_{\Gamma}$ posterior spans nearly the entire prior range. Because the true $\Gamma$ distribution is narrower for the power-law population, there is less deviation in the shape of the individual-event $\Gamma$ posteriors, which are not very informative to begin with. This leads to more uncertainty in the peak of the distribution but a better measurement of the spread. The $\mu_{k}$ posterior does not peak at the true value because higher values of $k$ are strongly disfavored by the prior in the individual-event PE, but the true value is included in the $3\sigma$ region. When compared to the $\mu_{k}$ posterior for the top-hat population, this is clear evidence for a structured jet, even though the exact value of the power-law index is not recovered accurately. The $\sigma_{k}$ posterior is again consistent with 0, as expected for a delta function distribution. 

The PPDs for the power-law population are shown in Fig.~\ref{fig:ppd_samplek}. The distribution for $\theta_{j}$ is again the best recovered. Even though $\sigma_{\Gamma}$ and $\sigma_{\log{E_{0}}}$ are well constrained, the PPDs for $\Gamma$ and $\log{E_{0}}$ are very broad because of the large uncertainty in $\mu_{\Gamma}$ and $\mu_{\log{E_{0}}}$. The distribution for $k$ does not peak at the true value, but the top-hat model with $k=0$ is excluded at 90\% confidence for this population, again clearly indicating the presence of jet structure for these simulations.
\begin{figure}
\centering
\centerline{\includegraphics[width=0.5\textwidth]{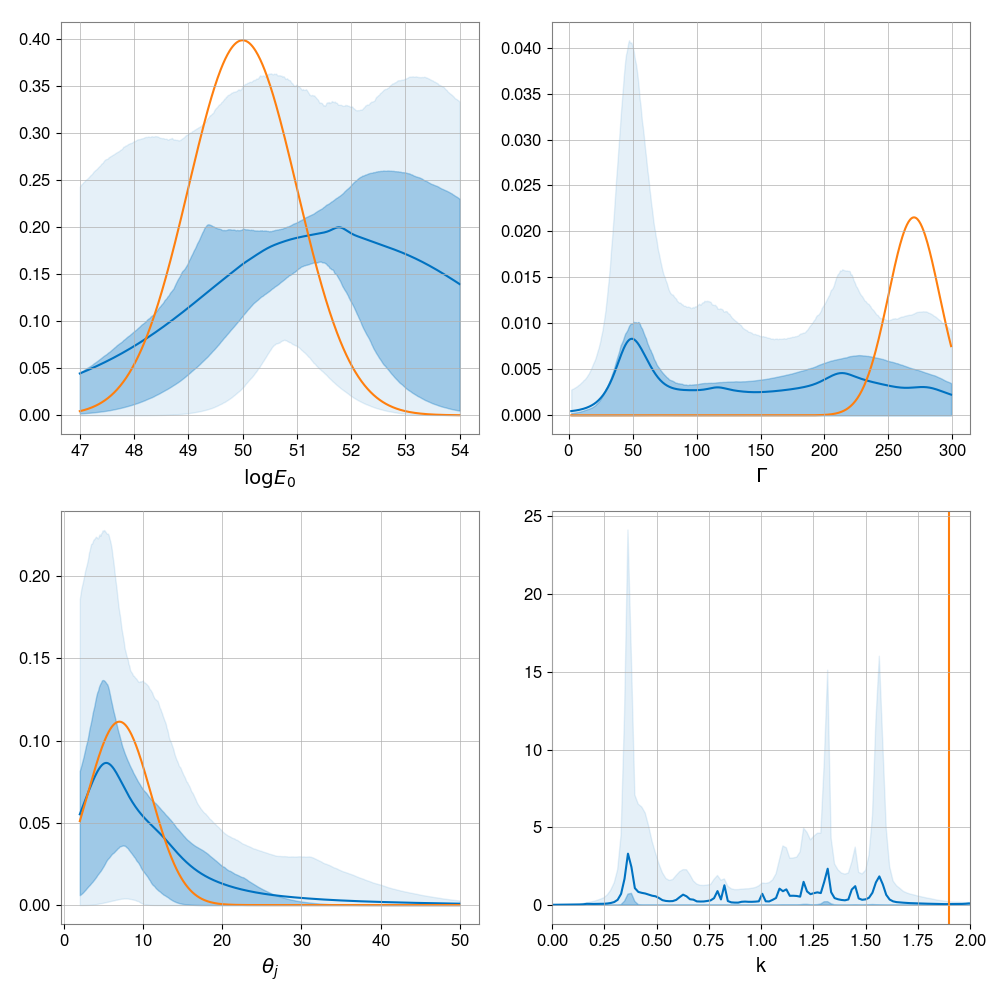}}
\caption{Posterior predictive distributions for the EM-only parameters for the power-law population set (dark blue lines) along with the 50\% and 90\% confidence intervals and the true distributions (orange lines).}
\label{fig:ppd_power_law}
\end{figure}

\section{Discussion}
\label{sec:discussion}

In this paper, we have developed a new method for determining the energy, Lorentz factor, opening angle, and power-law index for individual short GRBs as well as the distributions of these parameters for a given population of detected binary neutron star gravitational-wave events with an associated gamma-ray burst observation or fluence upper limit. Our method is completely independent of afterglow observations and uses Bayesian inference to combine the information provided by gravitational-wave parameter estimation on the inclination angle and distance to the source with the fluence measured by GRB satellites. 
We have simulated two populations of short GRBs-- one with top-hat jet geometry, and another with a power-law structured jet geometry with $k=1.9$. For individual events, little information is obtained for the jet geometry parameters $\theta_{j},\ \Gamma,\ \mathrm{and}\ k$ because of the degeneracy of the parameter space, but the $\log{E_{0}}$ of the jet can be constrained with an uncertainty of $\sim 2.5~\mathrm{dex}$. The hyper-parameters that describe the population as a whole are better constrained by combining the information from all 100 events in our simulation. For both jet structures, the peak of the opening angle distribution can be measured to within $10^{\circ}$. The peak of the energy distribution is also relatively well-reconstructed with an uncertainty of $\sim 2~\mathrm{dex}$ for the top-hat population. The $\Gamma$ distribution is the most difficult to constrain, since informative posteriors on this parameter in the individual event analysis depend on the observer looking right at the jet edge. Because the top-hat model is a subset of the power-law model as we have defined it with $k=0$, the hyper-parameter posterior for $\mu_{k}$ allows for us to distinguish between the two jet structures. The $\mu_{k}$ distribution does not accurately recover the true value for the power-law jet simulation because values of $k\gtrsim 1$ are strongly disfavored by the prior in the individual-event PE, but it does provide clear evidence for the universal structured jet model, ruling out a power law index of $k=0$ at 90\% confidence. 

The method we have developed offers the unique advantage of being independent of observations of the multi-wavelength GRB afterglow that traditional measurements of GRB jet parameters rely on. Even though the current alert system allows for the rapid follow-up of gravitational-wave BNS merger candidates by X-ray, optical, and radio telescopes, the sky localization is often prohibitively large for telescopes with small fields of view~\citep{LVCEMGuide, observing_scenarios}. Our method will provide increased statistics since the GRB satellite data can be searched a-posteriori for a coincident detection or upper limit. This also holds as the sensitivity of gravitational-wave detectors improves and the number of BNS candidate events increases, since electromagnetic partners will have to prioritize which events to follow up. 

Measurements of the energy, opening angle, Lorentz factor, and power-law index distributions of the short GRB population will have a significant impact on the theory of how these jets are launched and evolve. Constraints on the energy distribution to the level that we have demonstrated will enable distinguishing if the range of isotropic equivalent energies and luminosities of observed sGRBs is due to differences in the intrinsic brightness of the jet or if it is an artifact of observing the emission at different inclination angles. Measurements of the opening angle distribution could be used to determine the efficiency of jet collimation and could reveal two distinct populations of jets. Successful jets that manage to drill through, and eventually break out of, the merger ejecta will have narrower opening angles and smaller power-law indexes. On the other hand, ``choked" jets that cannot escape the merger ejecta will still inflate cocoons, but the resulting gamma-ray emission will appear to have a wider opening angle and a much lower Lorentz factor, which is an effect that we have shown can be measured. Together with the Lorentz factor distribution and power-law index, the opening angle distribution can also provide information on the density of circum-burst environment, since more interaction between the jet material and merger ejecta leads to slower and wider jets.

We conclude by considering some caveats to our analysis. The first is that neither of the two populations of GRBs that we have simulated matches the observed fluence distribution~\citep{fermi_catalog3}. This is in part due to the fact that the observed distribution is convolved with the instrument selection function, which we have not considered here since we run our analysis even on undetectable GRBs for which only upper limits on the fluence would be available. For this analysis, we also ignore the cosmological ``k-correction" that should be applied to account for the limited bandwidth of the GRB satellite and the effect of redshift on the observed photon frequency~\citep{Bloom_2001}. However, for the small distances we consider here, the effects of redshift are negligible and we argue that the method developed in Section~\ref{sec:energy_models} in terms of the bolometric fluence holds in this regime. Selection effects aside, the discrepancy in the fluence distributions indicates that the distributions we have chosen for the EM-only parameters do not correspond to the actual astrophysical distributions. Since the goal of our method is to measure these distributions, we had to make some assumptions for the purposes of our simulation. We chose to use Gaussian distributions since they are straightforward to parameterize, but our method could be extended to other distributions with different hyper-parameters.

We also only consider two angular profiles, the top-hat and power-law universal structured jet, both of which are simplified models that do not encompass the full details of the evolution of the jet emission. We emphasize that this analysis is a proof of principle and that our method could be extended to use models that parameterize more astrophysically-motivated scenarios like Gaussian structured jets~\citep{gaussian_jet_Xie_2018} or the generic ``boosted fireball"~\citep{boosted_fireball_Wu_2018, boosted_fireball_Wu_2019} as these more realistic models are starting to become available. While currently our method requires the jet structure model to be chosen before running the analysis, it could be applied to multiple jet structures to conduct Bayesian model selection between them at the individual-event or population level. We leave this to future studies.

By only including BNS progenitors out to 80 Mpc in our simulations, we have avoided the problem of gravitational-wave selection effects since all sources will be detectable out to this distance once the detectors reach design sensitivity (and we argue that even if a detection threshold were applied on the gravitational-wave data, it would not affect the inference of our chosen electromagnetic hyper-parameters in Appendix~\ref{sec:AppendixBayesian}). The SNR of the GW signal has very little impact on the shape of the individual event posteriors shown in Section~\ref{sec:individual}, which are instead dominated by the SNR of the GRB signal. We therefore expect that the results obtained here would also hold for a population of 100 detectable BNS sources even out to farther distances. Assuming even the most optimistic BNS merger rate, it would take much longer than the proposed lifetime of second-generation gravitational-wave detectors to reach 100 detections of BNSs within 80~Mpc. However, we could reach 100 total BNS detections out to farther distances by the end of advanced LIGO's fifth observing run (O5) with a five-detector network operating at a BNS detection range of at least $\sim 200~\mathrm{Mpc}$ for a realistic merger rate~\citep{gwtc1, observing_scenarios}. A fluence measurement or upper limit will not be available for every GRB associated with a detectable BNS, since the current GRB satellite network consisting of \textit{Fermi}-GBM, \textit{Swift}-BAT, \textit{INTEGRAL}, and the \textit{Interplanetary Network} (IPN) has an all-sky duty cycle of $\sim65\%$~\citep{Ajello_2019, Howell_2019}. Reaching the 100 gravitational-wave events with a GRB detection or fluence upper limit we have simulated here could be accomplished by the end of O5 if the detectors are running at the upgraded A+ sensitivity with a BNS detection range of $330~\mathrm{Mpc}$~\citep{white_paper} for a more optimistic merger rate of $\sim 2000~\mathrm{Gpc}^{-3}\ \mathrm{yr}^{-1}$ assuming one year of observation~\citep{gwtc1}, and is definitely achievable at LIGO Voyager sensitivity--a proposed upgrade to the existing advanced LIGO facilities in the late 2020s--which has a projected BNS range out to 1100 Mpc~\citep{white_paper, voyager}. Because we use both joint detections and non-detections with fluence upper limits in our analysis to constrain the population hyper-parameters, the rates we quote here are more optimistic than the joint GW-GRB detection rates in other work~\citep{Howell_2019}. The addition of next-generation GRB instruments like \emph{THESEUS}~\citep{theseus1, theseus2}, \emph{BurstCube}~\citep{burst_cube}, and \emph{HERMES}~\citep{hermes} coincident with the gravitational-wave detector upgrades to Voyager or even A+ sensitivity will greatly increase the all-sky duty cycle of the GRB satellite network and the joint detection rate. We note that based on the results in Figs.~\ref{fig:samplek_hyper_parameters} and \ref{fig:power_law_hyper_parameters}, the opening angle, power-law index, and energy distributions are well-constrained even with 40 coincident events, which is achievable with second-generation detectors for realistic merger rates. 
\acknowledgments
\textit{Acknowledgments.} S.B. and S.V. acknowledge support of the National Science Foundation and the LIGO Laboratory. LIGO was constructed by the California Institute of Technology and Massachusetts Institute of Technology with funding from the National Science Foundation and operates under cooperative agreement PHY-1764464.
S.B. is also supported by the American-Australian Fulbright Commission Postgraduate Research Fellowship, the Paul and Daisy Soros Fellowship for New Americans, and the NSF Graduate Research Fellowship under Grant No. DGE-1122374. ET is supported by ARC CE170100004 and ARC FT150100281.
The authors would like to thank Tsvi Piran, Ore Gottlieb, and Colm Talbot for helpful discussions, and Katerina Chatziioannou, Kentaro Mogushi, and Eric Howell for useful comments on the manuscript.
The authors acknowledge the LIGO Data Grid clusters through NSF Grant PHY-1700765 and the Ozstar supercomputing grid. LIGO Document Number P1900330.

\appendix
\section{Evaluation of the fluence integral}\label{sec:AppendixIntegral}

The integral for the fluence in Eq.~\ref{eq:fluence} is costly to evaluate, and must be computed a prohibitively large number of times while sampling. To contain the cost, we calculated it numerically using a Riemann sum with 1000 bins in $\theta$ between 0 and $\theta_{j}$ and 1000 bins in $\phi$ between 0 and $2\pi$. The core angle $\theta_{c}$ is chosen to be $1^{\circ}$ for power law jets with $k>0$. Because even the numerical integration would be prohibitively time-consuming when evaluating the fluence for each prior sample, a lookup table is constructed by evaluating the integral on a 4-dimensional grid in Lorentz factor, opening angle, power-law index, and inclination angle. The grid spacing for each parameter is detailed in Table~\ref{tab:integral_grid}. Points on the grid that violate the condition set in Eq.~\ref{eq:kmax} and thus represent unphysical parts of parameter space are left blank. When the fluence is evaluated for each prior sample, the value of the integral is then obtained via a nearest-neighbor interpolation using the lookup table. We have verified that discretizing the parameter space in this way does not impact the results.

\begin{table}[htb]
    \centering
\begin{tabular}{|p{1cm} ||p{1cm} p{1cm} p{1cm}|}
    \hline
     & min & max & $\Delta$ \\
    \hline \hline
    $\Gamma$ & 2 & 299 & 3 \\
    $\theta_{j}$ & $2^{\circ}$ & $50^{\circ}$ & $1^{\circ}$ \\
    k & 0 & 8 & 0.1 \\
    $\iota$ & $0^{\circ}$ & $90^{\circ}$ & $1^{\circ}$\\
    \hline
\end{tabular}
\caption{Minimum and maximum values and grid spacing for the 4-dimensional grid constructed for interpolating the value of the integral in the fluence expression, Eq.~\ref{eq:fluence}.}
\label{tab:integral_grid}
\end{table}

\section{Details of the Bayesian analysis implementation}\label{sec:AppendixBayesian}
\subsection{Individual-event analysis}
While Eq.~\ref{eq:posterior} is valid for constructing the posterior for continuous parameters, we obtain a series of discrete posterior samples for each of gravitational-wave parameters, so the value of the marginalized GW likelihood $\mathcal{L}^{\mathrm{GW}}(h |\boldsymbol{\eta})$ is not known directly but can be extracted if
the prior and the gravitational-wave evidence,
\begin{align}
\mathcal{Z}_{\mathrm{GW}} = \int \mathcal{L}^{\mathrm{GW}}(h| \boldsymbol{\eta}, \boldsymbol{\lambda})\pi(\boldsymbol{\eta})\pi(\boldsymbol{\lambda})\ d\boldsymbol{\eta}\ d\boldsymbol{\lambda},
\end{align}
and priors, $\pi(\boldsymbol{\eta})$ and $\pi(\boldsymbol{\lambda})$, are known. $\mathcal{Z}_{\mathrm{GW}}$ is calculated by the sampler in the gravitational-wave parameter estimation step described above, so the likelihood marginalized over the parameters unique to the gravitational-wave analysis, $\boldsymbol{\lambda}$, can be rewritten as:
\begin{align}
\mathcal{L}^{\mathrm{GW}}(h\vert \boldsymbol{\eta}) = \frac{\mathcal{Z}_{\mathrm{GW}}p(\boldsymbol{\eta}\vert h)}{\pi(\boldsymbol{\eta})}.
\label{eq:gw_likelihood}
\end{align}
The parameter vector $\boldsymbol{\eta}$ is not continuous, but rather a list of $k$ n-tuples, where $n=4$ for this analysis. The probability of each $\boldsymbol{\eta}_{i}$ is $p(\boldsymbol{\eta}_{i}) = 1/k$. If we substitute the likelihood from Eq.~\ref{eq:gw_likelihood} into the joint likelihood function defined in Eq.~\ref{eq:joint_likelihood}, we obtain
\begin{align}
\mathcal{L}^{\mathrm{EM+GW}}(h,F^{\gamma}\vert \mathbf{x}) = \mathcal{Z}_{\mathrm{GW}}\int p(\boldsymbol{\eta}\vert h)\mathcal{L}^{\mathrm{EM}}(F^{\gamma} \vert \mathbf{x},\boldsymbol{\eta})\, d\boldsymbol{\eta},
\end{align}
which is just the expectation value of the EM likelihood
\begin{align}
\mathcal{L}^{\mathrm{EM+GW}}(h,F^{\gamma}\vert \mathbf{x}) = \mathcal{Z}_{\mathrm{GW}}\langle \mathcal{L}^{\mathrm{EM}}(F^{\gamma} \vert \mathbf{x},\boldsymbol{\eta}) \rangle,
\end{align}
since $p(\boldsymbol{\eta} | h)$ is a normalized probability distribution. For a discrete set of posterior samples, this expression becomes
\begin{align}
\mathcal{L}^{\mathrm{EM+GW}}(h,F^{\gamma}\vert \mathbf{x}) = \frac{\mathcal{Z}_{\mathrm{GW}}}{k}\sum_{i=1}^{k}\mathcal{L}^{\mathrm{EM}}(F^{\gamma} \vert \mathbf{x},\boldsymbol{\eta}_{i}),
\label{eq:discrete_joint_likelihood}
\end{align}
so the posteriors for the EM-only parameters $\mathbf{x}$ are obtained by ``recycling" the posteriors on the common parameters obtained from the gravitational-wave parameter estimation step~\citep{Abbott_2019, thrane_talbot_2019}:
\begin{align}
    p(\mathbf{x}|h, F^{\gamma}) = \frac{\pi(\mathbf{x})}{\mathcal{Z}_{\mathbf{x}}}\frac{\mathcal{Z}_{\mathrm{GW}}}{k}\sum_{i=1}^{k}\mathcal{L}^{\mathrm{EM}}(F^{\gamma} \vert \mathbf{x},\boldsymbol{\eta}_{i}).
\end{align}

\subsection{Hierarchical Modeling}
The likelihood for the hyper-parameters defined in Eq.~\ref{eq:hyper_likelihood} can be recast in terms of the posterior on the EM-only parameters $\mathbf{x}$ that we have already obtained in the previous sampling step:
\begin{align}
\mathcal{L}(h, F^{\gamma} \vert \boldsymbol{\Lambda}) &= \int d\mathbf{x} \mathcal{L}^{\mathrm{GW+EM}}(h, F^{\gamma} \vert \mathbf{x}, \boldsymbol{\Lambda})\pi(\mathbf{x} \vert \boldsymbol{\Lambda})\\
&= \int d\mathbf{x} \frac{p(\mathbf{x} \vert h, F^{\gamma})\mathcal{Z}_{\mathbf{x}}}{\pi_{0}(\mathbf{x})}\pi(\mathbf{x} \vert \boldsymbol{\Lambda}),
\end{align}
where $\mathcal{Z}_{\mathbf{x}}$ is the EM evidence defined in Eq.~\ref{eq:em_evidence} and $\pi_{0}(\mathbf{x})$ is the prior used in sampling the EM-only parameters first presented in Eq.~\ref{eq:posterior}. The likelihood for the hyper-parameters is the expectation value of the ratio of the hyper-prior to the original prior because the posterior $p(\mathbf{x} | h, F^{\gamma})$ is a normalized probability distribution function,
\begin{align}
\mathcal{L}(h, F^{\gamma} \vert \boldsymbol{\Lambda}) &=\mathcal{Z}_{\mathbf{x}}\bigg\langle \frac{\pi(\mathbf{x} \vert \boldsymbol{\Lambda})}{\pi_{0}(\mathbf{x})}\bigg\rangle,
\end{align}
which can be written as a sum over the posterior samples obtained for the EM-only parameters $\mathbf{x}_{i}$ in the second sampling step described above for an individual event:
\begin{align}
\mathcal{L}(h, F^{\gamma} \vert \boldsymbol{\Lambda}) &=\frac{\mathcal{Z}_{\mathbf{x}}}{n}\sum_{i}^{n}\frac{\pi(\mathbf{x}_{i} \vert \boldsymbol{\Lambda})}{\pi_{0}(\mathbf{x}_{i})}.
\end{align}

The hyper-parameter posterior for a population of $N$ events defined in Eq.~\ref{eq:hyper_posterior} can then be written in terms of the sum over samples as:

\begin{align}
p(\boldsymbol{\Lambda} \vert \{h\}, \{F^{\gamma}\})&= \frac{\pi(\boldsymbol{\Lambda})}{\mathcal{Z}_{\boldsymbol{\Lambda}}}\prod_{j}^{N}\mathcal{L}(h_{j}, F^{\gamma}_{j} \vert \boldsymbol{\Lambda})\\
&=\frac{\pi(\boldsymbol{\Lambda})}{\mathcal{Z}_{\boldsymbol{\Lambda}}}\prod_{j}^{N}\frac{\mathcal{Z}_{\mathbf{x}_j}}{n_{j}}\sum_{i}^{n_{j}}\frac{\mathcal{N}(\mathbf{x}_{ij}, \boldsymbol{\Lambda})}{\pi_{0}(\mathbf{x}_{ij})},
\end{align}
where we have substituted the truncated multivariate Gaussian $\mathcal{N}(\mathbf{x}_{ij}, \boldsymbol{\Lambda})$ for the hyper-prior $\pi(\mathbf{x}_{i} \vert \boldsymbol{\Lambda})$, and the hyper evidence, $\mathcal{Z}_{\boldsymbol{\Lambda}}$ is given by marginalizing the likelihood in Eq.~\ref{eq:hyper_likelihood}:
\begin{align}
    \mathcal{Z}_{\boldsymbol{\Lambda}} = \int d\boldsymbol{\Lambda}\ \pi(\boldsymbol{\Lambda}) \prod_{j}^{N}\mathcal{L}(h_{j}, F^{\gamma}_{j} \vert \boldsymbol{\Lambda}).
\end{align}
\subsection{Selection Effects}
Below we demonstrate that as long as a detection threshold based on the GRB parameters is never introduced, selection effects do not enter the method we've developed. Even if a detection threshold is imposed based on the gravitational wave parameters, selection biases do not affect our analysis because the population properties we seek to characterize are independent of the GW parameters. We follow the arguments presented in Appendix E1 of \cite{thrane_talbot_2019}. If we impose an arbitrary detection threshold on the gravitational-wave matched filter SNR, $\rho_{\mathrm{mf}} > \rho_{\min}$, where
\begin{align}
    \rho_{\mathrm{mf}} = \frac{\langle h, \hat{h}(\boldsymbol{\eta}, \boldsymbol{\lambda})\rangle}{\sqrt{\langle \hat{h}(\boldsymbol{\eta}, \boldsymbol{\lambda}), \hat{h}(\boldsymbol{\eta}, \boldsymbol{\lambda})\rangle}}
\end{align}
for the inner product defined as
\begin{align}
    \langle a, b\rangle = \frac{4}{T} \sum_{k} \Re\left(\frac{a^{*}(f_{k})b^{*}(f_{k})}{S_{h}(f_{k})}\right),
\end{align}
then the likelihood in Eq.~\ref{eq:gw_likelihood} needs to be modified so that it remains properly normalized with respect to the data, $h$:
\begin{align}
    \mathcal{L}^{\mathrm{GW}}(h | \boldsymbol{\eta}, \boldsymbol{\lambda}, \mathrm{det}) &=
    \begin{cases}
    \frac{1}{p_{\mathrm{det}}(\boldsymbol{\eta}, \boldsymbol{\lambda})}\mathcal{L}^{\mathrm{GW}}(h | \boldsymbol{\eta}, \boldsymbol{\lambda})\ &\rho_{\mathrm{mf}} \geq \rho_{\min}\\
    0\ &\rho_{\mathrm{mf}} < \rho_{\min}
    \end{cases}.
\end{align}
The detection probability, $p_{\mathrm{det}}(\boldsymbol{\eta}, \boldsymbol{\lambda})$ is defined as:
\begin{align}
    p_{\mathrm{det}}(\boldsymbol{\eta}, \boldsymbol{\lambda}) &= \int_{\rho_{\mathrm{mf}} > \rho_{\min}} dh\ \mathcal{L}^{\mathrm{GW}}(h | \boldsymbol{\eta}, \boldsymbol{\lambda}). \label{eq:pdet}
\end{align}
The prior on the GW parameters needs to be similarly modified to account for the preference of detecting sources in certain parts of the sky with certain masses, etc. which can be quantified \textit{a priori} using simulations:
\begin{align}
    \pi(\boldsymbol{\eta}, \boldsymbol{\lambda}| \mathrm{det}) = \frac{\pi(\boldsymbol{\eta}) \pi(\boldsymbol{\lambda})p_{\mathrm{det}}(\boldsymbol{\eta}, \boldsymbol{\lambda})}{\int \pi(\boldsymbol{\eta}) \pi(\boldsymbol{\lambda})p_{\mathrm{det}}(\boldsymbol{\eta}, \boldsymbol{\lambda})d\boldsymbol{\eta} d\boldsymbol{\lambda}}.
\end{align}
The denominator is a constant normalization factor, which is the same for all coincident events in our population as long as the same GW prior is used. 
For a detected event, the joint likelihood in Eq.~\ref{eq:joint_likelihood} is now
\begin{align}
     \mathcal{L}^{\mathrm{GW+EM}}(h,F^{\gamma}|\mathbf{x}, \mathrm{det}) &= \int \frac{\mathcal{L}^{\mathrm{GW}}(h | \boldsymbol{\eta}, \boldsymbol{\lambda})}{p_{\mathrm{det}}(\boldsymbol{\eta}, \boldsymbol{\lambda})} \mathcal{L}^{\mathrm{EM}}(F^{\gamma} | \mathbf{x},\boldsymbol{\eta})\frac{\pi(\boldsymbol{\eta}) \pi(\boldsymbol{\lambda})p_{\mathrm{det}}(\boldsymbol{\eta}, \boldsymbol{\lambda})}{\int \pi(\boldsymbol{\eta}) \pi(\boldsymbol{\lambda})p_{\mathrm{det}}(\boldsymbol{\eta}, \boldsymbol{\lambda})d\boldsymbol{\eta} d\boldsymbol{\lambda}} d\boldsymbol{\eta} d\boldsymbol{\lambda} \\
     &= \frac{1}{C_{\mathrm{det}}}\int \mathcal{L}^{\mathrm{GW}}(h | \boldsymbol{\eta}, \boldsymbol{\lambda})\mathcal{L}^{\mathrm{EM}}(F^{\gamma} | \mathbf{x},\boldsymbol{\eta})\pi(\boldsymbol{\eta})\pi(\boldsymbol{\lambda}) d\boldsymbol{\eta}\ d\boldsymbol{\lambda}\\
     &= \frac{1}{C_{\mathrm{det}}} \mathcal{L}^{\mathrm{GW+EM}}(h, F^{\gamma} | \mathbf{x})
\end{align}
where we have defined 
\begin{align}
    C_{\mathrm{det}} = \int \pi(\boldsymbol{\eta}) \pi(\boldsymbol{\lambda})p_{\mathrm{det}}(\boldsymbol{\eta}, \boldsymbol{\lambda})d\boldsymbol{\eta} d\boldsymbol{\lambda}.
\end{align}
Because $C_{\mathrm{det}}$ is a constant for all events in the population, this extra normalization factor can be practically ignored and the method described in Section~\ref{sec:methods} is unaffected by the detection threshold introduced on the gravitational-wave signal. We stress that this is only true if the population parameters are independent of the GW parameters, i.e.:
\begin{align}
    \pi(\boldsymbol{\eta}, \boldsymbol{\lambda}, \mathbf{x} | \boldsymbol{\Lambda}) = \pi(\mathbf{x} | \boldsymbol{\Lambda}) \pi(\boldsymbol{\eta})\pi(\boldsymbol{\lambda}),
\end{align}
and if there is no cut applied on the GRB fluence in order for the event to be included in our analysis. We have enforced this in our analysis by randomly assigning GRB parameters drawn from the hyper-prior $\pi(\mathbf{x} | \boldsymbol{\Lambda})$ to each gravitational-wave source and by running on GRBs with arbitrarily low fluences.

\bibliographystyle{apj}
\bibliography{sgrb}

\end{document}